\documentclass[%
 twocolumn,amsmath,amssymb,aps, nofootinbib
]{revtex4-2}

\usepackage{graphicx}
\usepackage[dvipsnames]{xcolor}
\usepackage{dcolumn}% Align table columns on decimal point
\usepackage{bm}% bold math

\usepackage[colorlinks=true,
    linkcolor=blue,
    filecolor=blue,
    citecolor=blue,      
    urlcolor=blue,]{hyperref}% add hypertext capabilities
\usepackage{subfig}
\usepackage{mhchem}
\usepackage{multirow}

\usepackage{prettyref}
\newcommand{\pref}[1]{\prettyref{#1}}
\newrefformat{fig}{Fig.~\ref{#1}}
\newrefformat{tab}{Table~\ref{#1}}
\newrefformat{sec}{Sec.~\ref{#1}}
\newrefformat{subsec}{Sec.~\ref{#1}}
\newrefformat{eq}{Eq.~(\ref{#1})}

\begin{document}

\title{Quadrupole formation and coupling to magnetic and structural degrees of freedom in the $5d^1$ double perovskites Ba$_2$MgReO$_6$ and Ba$_2$NaOsO$_6$}

\author{Francesco Martinelli}
\author{Claude Ederer}
\affiliation{Materials Theory, ETH Z\"{u}rich, Wolfgang-Pauli-Strasse 27, 8093 Z\"{u}rich, Switzerland}

\date{\today}

\begin{abstract}
We investigate the interplay between charge, magnetic, and structural degrees of freedom in the isostructural and isoelectronic $5d^1$ double-perovskites Ba$_2$MgReO$_6$ and Ba$_2$NaOsO$_6$. Using first-principles-based electronic structure calculations, we show that both materials exhibit a tendency toward spontaneous quadrupolar order in the cubic paramagnetic phase, which is slightly weaker in Ba$_2$NaOsO$_6$ than in Ba$_2$MgReO$_6$. Our analysis further reveals an intimate coupling between the local magnetic moments and charge quadrupoles, mediated by the strong spin-orbit interaction, that leads to the unusual canted configuration of magnetic moments observed in these systems. When structural degrees of freedom are included, the two materials exhibit pronounced differences. In Ba$_2$MgReO$_6$ the strong coupling to Jahn-Teller distortions stabilizes the antiferroic $\mathcal{Q}_{x^2-y^2}$ order, yielding excellent agreement with available experimental data. In contrast, the Jahn-Teller coupling is significantly weaker in Ba$_2$NaOsO$_6$ and appears insufficient to stabilize the antiferroic quadrupolar order. While this is consistent with the absence of any measurable long-range structural distortion above the magnetic transition temperature, it contrasts with experimental results indicating a strong canting of the magnetic moments. Our analysis thus successfully describes the mechanisms shaping the properties of the Re-compound while a full quantitative description of the magnetic ground state of Ba$_2$NaOsO$_6$ is still elusive.
\end{abstract}

\maketitle

%%%%%%%%%%%%%%%%%%%%%%%%%%%%%%%%%%%%%%%%%%%%%%%%%%%%%%%%%%%%%%%%%%%%%%%%%%%%%%%%

\section{Introduction}

The exploration of exotic states of matter, such as, e.g., high-temperature superconductivity~\cite{Dagotto:1994}, quantum spin liquids~\cite{Savary/Balents:2016}, or Mott-insulating states~\cite{Imada/Fujimori/Tokura:1998}, and how such phases emerge from the collective interplay between electronic, magnetic, orbital, and structural degrees of freedom, is a central topic of modern condensed matter physics. Recently, certain complex oxides containing either $4d$ or $5d$ transition metal cations have entered the spotlight. In these materials, the interplay of strong spin-orbit coupling and electronic correlations can give rise to intriguing multipolar orderings, exotic spin textures, and non-trivial topological phases~\cite{Witczak-Krempa_et_al:2014, Rau/Lee/Kee:2016, Martins/Aichhorn/Biermann:2017, Takayama_et_al:2021}.

\begin{figure}
    \centering
    \includegraphics[width=0.4\textwidth]{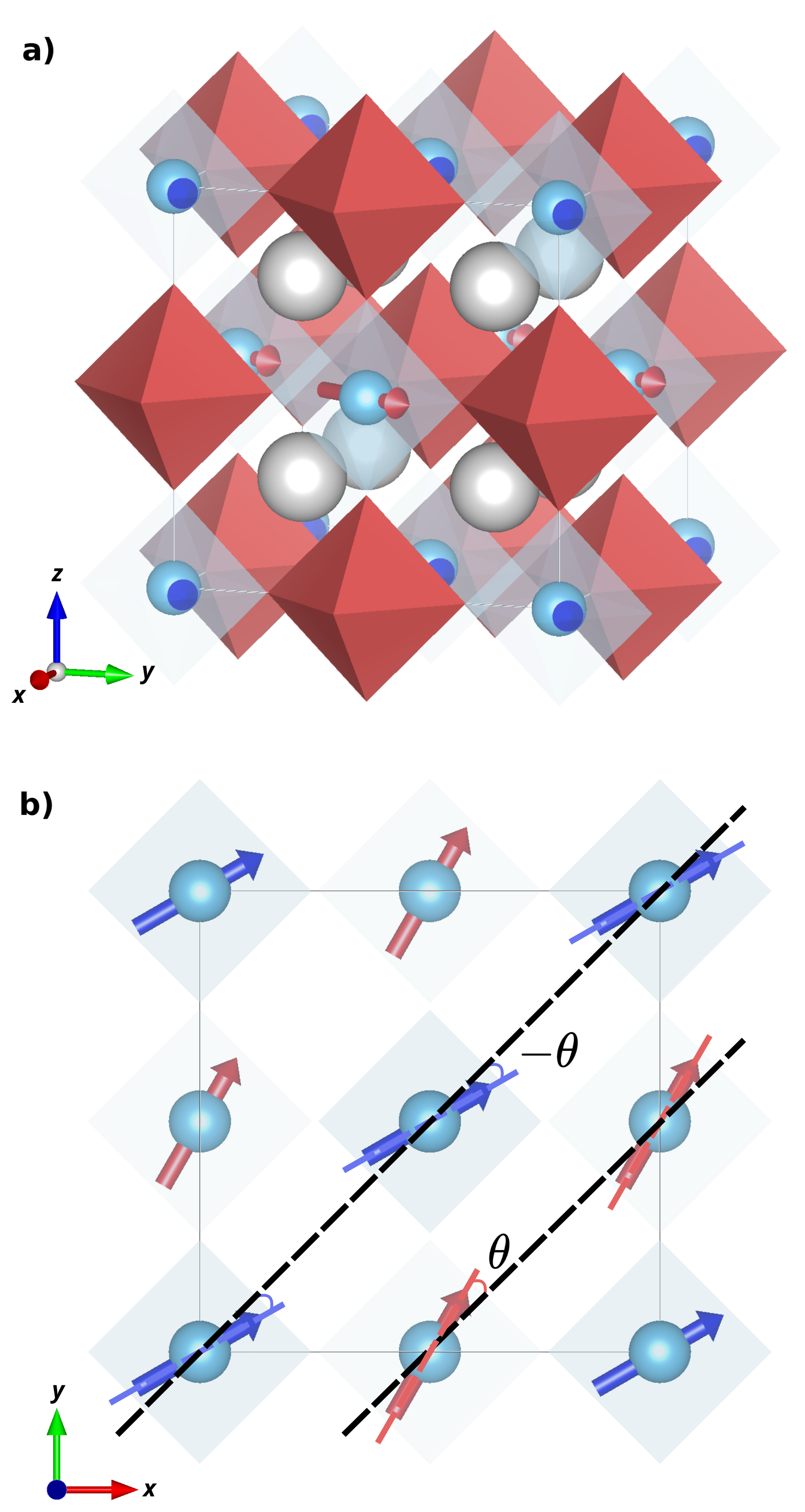}
    \caption{(a) Three-dimensional view of the double perovskite structure ($A_2BB'$O$_6$) showing the canted ferromagnetic configuration. $A$ ions are shown in gray and transition metal ions ($B'$) in light blue, with arrows indicating the local magnetic moments. The oxygen octahedra surrounding the $B$ and $B'$ ions are indicated in red and transparent gray. (b) The local magnetic moments are rotated away from the [110] direction (indicated by the black dashed lines) by an angle $\theta$, with $\theta>0$ indicating counterclockwise rotation around $z$. The thin gray line indicates the conventional cubic unit cell. Note that the Re atoms indicated with blue and red dipole moments are located in different adjacent [001] planes [see also subfigure (a)].}
    \label{fig:bmrocFM_new}
\end{figure}

Particularly interesting are $B$-site-ordered double-perovskite oxides, with the general chemical composition $A_2BB'$O$_6$, where $A$ and $B$ are nonmagnetic closed shell ions and $B'$ is a magnetic $4d$ or $5d$ transition metal cation. 
The high chemical flexibility of these double perovskites, which allows to incorporate many different combinations of ions on the $A$, $B$, and $B'$ sites~\cite{Vasala/Karppinen:2015}, in combination with their simple high symmetry crystal structure [see \pref{fig:bmrocFM_new}(a)], facilitates systematic studies over closely related series of compounds. 
Indeed, double perovskites have been found to host a variety of exotic quantum states, including symmetry-broken phases where the order parameter is a higher-order multipole of the charge or magnetization density as well as potential spin liquid or glassy states with no detectable symmetry breaking down to very low temperatures~\cite{Chen/Pereira/Balents:2010, Ishizuka/Balents:2014, Romhanyi/Balents/Jackeli:2017, Lu_et_al:2017, Liu_et_al:2018a, Willa_et_al:2019, Hirai/Hiroi:2019, Hirai_et_al:2020, Maharaj_et_al:2020a, Iwahara/Chibotaru:2023, Soh_et_al:2024, Agrestini_et_al:2024, Zivkovic_et_al:2024}. 

The complexity of the multipolar physics in these materials, involving strong entanglement between spin, orbital, and structural degrees of freedom, poses a significant challenge for the theoretical modeling.
Simultaneously, higher-order multipoles are not easily accessible using standard experimental probes, and instead require the application of advanced techniques and often several complementary measurements. Quantitatively predictive first-principles-based calculations are therefore extremely valuable to reveal a possible ordering of higher-order multipoles and elucidate the physical and chemical mechanisms that control their behavior \cite{Suzuki/Ikeda/Oppeneer:2018, FioreMosca_et_al:2021,  MansouriTehrani/Spaldin:2021, Urru_et_al:2023a, Verbeek/Urru/Spaldin:2023, Merkel/Tehrani/Ederer:2024, Otsuki_et_al:2024, FioreMosca/Franchini/Pourovskii:2024}.

Here, we focus on the two closely related double-perovskites, Ba$_2$MgReO$_6$ and Ba$_2$NaOsO$_6$, containing Re$^{6+}$ and Os$^{7+}$ cations on the $B'$ sites, respectively. Both cations exhibit a nominal $5d^1$ electron configuration, and the octahedral crystal field together with the strong spin-orbit coupling results in a four-fold degenerate $J_\text{eff}=3/2$ ionic ground state configuration~\cite{Takayama_et_al:2021}. This degeneracy of their ground state multiplet makes these systems sensitive to small distortions and prone to spontaneous symmetry breaking. Furthermore, the local magnetic dipole moment is expected to be strongly reduced due to partial cancellation between spin and orbital contributions, and the frustration of the nearest neighbor interaction on the face-centered cubic lattice formed by the $B'$ cations can lead to a further suppression of conventional antiferromagnetic dipolar order, potentially favoring the emergence of higher order multipoles. 

Anomalies in the heat capacity of Ba$_2$MgReO$_6$ at 33~K and 18~K \cite{Marjerrison_et_al:2016, Hirai/Hiroi:2019} suggest that this system undergoes two successive phase transitions, and recent experiments employing both resonant and off-resonant x-ray diffraction revealed that at $T_q = 33$~K the space group symmetry is reduced from $Fm\bar{3}m$ to $P4_2/mnm$. This symmetry lowering has been related to an antiferroic ordering of local $x^2-y^2$-type quadrupoles on the Re sites in combination with a ferroic $3z^2-r^2$-type quadrupolar ordering and a small Jahn-Teller distortion of the ReO$_6$ octahedra~\cite{Hirai_et_al:2020, Soh_et_al:2024}. The transition at $T_m=18$~K then corresponds to a non-collinear dipolar magnetic ordering with a net magnetization within a single domain of approximately 0.46~$\mu_B$ per Re along the [110] direction, but with the local magnetic moments canted away from the [110] direction by an angle $\theta$ of about $\pm$40$^\circ$ [see \pref{fig:bmrocFM_new}(b)], corresponding to an antiferromagnetic component of the order parameter with wave vector along [001]~\cite{Hirai/Hiroi:2019, Hirai_et_al:2020}.

In contrast, the case of Ba$_2$NaOsO$_6$ appears more complex. Below 6.8~K, this material exhibits a canted magnetic dipolar order similar to Ba$_2$MgReO$_6$, with a canting angle $\theta$ of $\pm 67^\circ$ and a net magnetization of 0.2 $\mu_B$ along the [110] direction~\cite{Erickson_et_al:2007, Lu_et_al:2017}. However, no deviation from global cubic symmetry or other evidence for long range quadrupolar order has been found above the magnetic transition temperature~\cite{Erickson_et_al:2007, daCruzPinhaBarbosa_et_al:2022}. Instead, a few K above $T_m$, the system exhibits signs for a broken local point symmetry, deduced from a splitting of the nuclear magnetic resonance (NMR) spectra, which indicates the emergence of a local electric field gradient around the Na ions, stemming from a local distortion or charge density anisotropy~\cite{Lu_et_al:2017, Liu_et_al:2018, Liu_et_al:2018a, Cong_et_al:2019, Willa_et_al:2019}.
Recently, it has been proposed that the broken local point symmetry in Ba$_2$NaOsO$_6$ is even retained up to at least room temperature, suggesting a dynamical Jahn-Teller scenario without global symmetry breaking~\cite{Agrestini_et_al:2024}. A dynamical Jahn-Teller effect persisting up to temperatures well above $T_q$ has also been suggested for Ba$_2$MgReO$_6$~\cite{Zivkovic_et_al:2024}.

Thus, despite the fact that Ba$_2$MgReO$_6$ and Ba$_2$NaOsO$_6$ are essentially isostructural and isoelectronic, subtle differences between the two materials seem to affect the possible emergence of a long-range ordered quadrupolar phase.

Several aspects of Ba$_2$MgReO$_6$ and Ba$_2$NaOsO$_6$ have been addressed previously using first-principles electronic-structure calculations based on density functional theory (DFT) and different extensions of DFT. For Ba$_2$NaOsO$_6$, most studies have examined the interplay between spin-orbit coupling and electronic correlations to clarify the nature of its insulating state~\cite{Xiang/Whangbo:2007, Gangopadhyay/Pickett:2015, FioreMosca_et_al:2024}, while Ref.~\cite{FioreMosca_et_al:2021} has demonstrated the relevance of multipolar spin interactions and a potential Jahn-Teller distortion for the reported canted magnetic order.
Similarly for the case of Ba$_2$MgReO$_6$, previous studies have focused on the influence of spin-orbit coupling on its Mott-insulating character, the emergence of quadrupolar order, and its coupling to magnetic and structural degrees of freedom~\cite{MansouriTehrani/Spaldin:2021, Merkel/Tehrani/Ederer:2024, Soh_et_al:2024, FioreMosca/Franchini/Pourovskii:2024}. In particular, these studies have indicated that a coupling between the quadrupolar order and the structural Jahn-Teller distortion is required to stabilize the experimentally observed antiferroic $x^2-y^2$-type order, while, in the absence of such coupling, a ferroic order of $t_{2g}$-type quadrupoles would be preferred~\cite{Soh_et_al:2024, FioreMosca/Franchini/Pourovskii:2024}. Furthermore, the microscopic model of coupled quadrupolar, dipolar, and vibronic modes presented in  Ref.~\cite{FioreMosca/Franchini/Pourovskii:2024} has successfully described the two successive phase transitions at low temperatures.

In the present work, we present a complementary first principles study, confirming the influence of the Jahn-Teller coupling on the stabilization of the experimentally observed antiferroic quadrupolar order in Ba$_2$MgReO$_6$. Furthermore, we provide a systematic comparison between Ba$_2$MgReO$_6$ and Ba$_2$NaOsO$_6$, aiming to identify the origin for their different behavior. Thereby, we focus specifically on clearly disentangling the coupling between the various degrees of freedom, i.e., electronic quadrupole moments, magnetic dipole moments, and different Jahn-Teller distortions. 
We find that Ba$_2$NaOsO$_6$ generally exhibits a slightly weaker tendency towards quadrupolar order, and that in particular the coupling to the Jahn-Teller modes is significantly weaker than in Ba$_2$MgReO$_6$, and is not able to stabilize the antiferroic quadrupolar order. We also demonstrate an intimate coupling between the direction of the local magnetic moments of the 5$d$ transition metal cations and the corresponding local quadrupole, and show that the canting of the magnetic moments away from the ferromagnetic [110] direction is driven by the emergence of an antiferroic component of the local quadrupoles (or vice versa), which rotates the local easy axis away from the global magnetization direction. 

Overall, we obtain excellent agreement with available experimental data for Ba$_2$MgReO$_6$. For Ba$_2$NaOsO$_6$, the weaker tendency towards quadrupolar order and the weak coupling to a corresponding Jahn-Teller distortion found in our calculations appears consistent with the absence of any long-range ordered quadrupolar and Jahn-Teller distorted phase above the magnetic ordering temperature. On the other hand, based on our analysis of the correspondence between antiferroic quadrupolar order and the canting of the local magnetic moments, the lack of stabilization of such an antiferroic quadrupolar order in our calculations appears to contradict the reported canting of the magnetic moments away from the [110] direction, based on the analysis of NMR spectra~\cite{Lu_et_al:2017}. Further research is needed to resolve this remaining discrepancy.

\section{Methods and Computational Details}

\subsection{Constraining quadrupole moments and magnetic dipoles}
\label{sec:method}

The quadrupole moment corresponding to an electron density $n(\vec{r})$ around a specific site $I$ can be defined as \mbox{$\mathcal{Q}^{I}_{ij}=\int d^3r\,  r_i\, r_j\, n(\vec{r})$}, with $r_i$ a cartesian component of the distance vector relative to the chosen site. Since $\mathcal{Q}_{ij}^{I}$ is by definition a totally symmetric tensor, it can  be decomposed into its trace and a symmetric traceless part, which form one- and five-dimensional irreducible representations under arbitrary rotations, respectively. The components of the latter can be labeled as  $\mathcal{Q}_t^{I}$, with $t \in \{xy, yz, 3z^2-r^2, xz,x^2-y^2\}$. 

Expressing the electron density in terms of the density matrix with respect to a suitably chosen atomic-orbital-like basis (labeled by the usual angular momentum and spin quantum numbers, $l$, $m$, and $s$), allows to write the components $\mathcal{Q}_t^{I}$ as follows:
\begin{align}\label{eq:0}
    \mathcal{Q}^I_t = \sum_{lm,l'm'} \mu^{2,t}_{lm,l'm'} \, \rho^{I}_{l'm',lm} \quad ,
\end{align}
where we use the same notation as in Ref.~\cite{Schaufelberger_et_al:2023} to express the matrix elements of the charge-multipole operators with rank $k$, $\mu^{k,t}_{lm,l'm'}$ \cite{Santini_et_al:2009, Bultmark_et_al:2009}, and consider only the angular part of the density matrix, $\rho^{I}_{l'm',lm} = \sum_{s}\rho^{I}_{l'm's,lms}$. 
Since for the materials considered here, the dominant contribution to the charge quadrupoles arises from the $d$-orbitals, we further restrict our analysis to $l=l'=2$.

To constrain the local charge quadrupoles, we use the method introduced in Ref.~\cite{Schaufelberger_et_al:2023}. Thereby, the energy as function of local quadrupoles is obtained by adding a linear form of the constraint to the DFT (or DFT+$U$) energy using the method of Lagrange multipliers, and then minimizing the resulting functional with respect to the density and the Lagrange multiplier: 
\begin{align}\label{eq:1}
& E[\{ \tilde{\mathcal{Q}}_t\}] = \min_{\bm{\rho}(\vec{r}), s^I_t} \Bigg( E_\text{0}[\bm{\rho}(\vec{r})] - \sum_{I,t} s^{I}_t \Big( \mathcal{Q}^I_t[\bm{\rho}(\vec{r})] - \tilde{\mathcal{Q}}^{I}_t \Big) \Bigg) .
\end{align}
Here, $\tilde{\mathcal{Q}}^I_t$ denote the specific values to which the quadrupoles are constrained, $s^I_t$ are the corresponding Lagrange multipliers, and $\bm{\rho}(\vec{r}) = n(\vec{r}) + \vec{\bm{\sigma}}\cdot \vec{m}(\vec{r})$ is the full spin-density, which describes both the electronic charge  and spin density of the system, with $\vec{\bm{\sigma}}$ representing the vector of Pauli matrices. 
The constraint results in an orbital-dependent local potential shift added to the usual Kohn-Sham potential, which induces a deformation of the electron density around the atom according to the corresponding multipole component. 
To obtain a specific value, $\tilde{\mathcal{Q}}^I_t$, the corresponding Lagrange multiplier, $s^I_t$, which represents the amplitude of the local potential shift, has to be adjusted self-consistently. Alternatively, one can drop this minimization with respect to $s^I_t$ and simply vary the strength of the potential shift within a suitable range to obtain properties as a function of the local quadrupole moments~\cite{Schaufelberger_et_al:2023}.

Apart from constraining the local quadrupoles, we also perform calculations where we constrain the directions of the local spin moments, in order to monitor the energy and other properties as function of the canting angle $\theta$ [see \pref{fig:bmrocFM_new}(b)] and also to simulate a disordered paramagnetic state.
Here, we use the well-established method of adding a penalty term to the DFT+$U$ energy, which corresponds to adding a quadratic form of the constraint to the energy functional~\cite{Ma/Dudarev:2015}:
\begin{equation}
\label{eq:m-constraint}
E=E_0[\bm{\rho}(\vec{r})] + \sum_I \lambda\left[\vec{S}_I-\hat{S}_I^0\left(\hat{S}_I^0 \cdot \vec{S}_I\right)\right]^2
\end{equation}
Here,  $\vec{S}_I = \vec{S}_I[\bm{\rho}(\vec{r})]$ is the local spin moment at site $I$ and $\hat{S}^0_I$ is a unit vector pointing along the direction along which $\vec{S}_I$ is constrained.
The parameter $\lambda$ in \pref{eq:m-constraint} determines the strength of the constraint, and has to be chosen such that corresponding penalty energy remains negligible compared to the energy scale of interest.

\subsection{Computational details}
\label{sec:comp-details}

We perform DFT+$U$ calculations, with spin-orbit coupling always included, within the projector-augmented wave (PAW) framework using the ``Vienna Ab-initio Simulation Package'' (VASP)~\cite{Hafner/Kresse:1997, Kresse/Joubert:1999} based on the exchange correlation energy functional of Perdew-Burke-Ernzerhof (PBE)~\cite{Perdew/Burke/Ernzerhof:1996}. We use the simplified rotational invariant formulation of DFT+$U$ first proposed by Dudarev et al.~\cite{Dudarev_et_al:1998}, where the local interaction is parametrized by a single parameter $U_{\text{eff}}$.

Most of our calculations are performed for the perfectly cubic structure with $Fm\bar{3}m$ space group symmetry using experimental structural data obtained at room temperature in Ref.~\cite{Hirai_et_al:2020} for Ba$_2$MgReO$_6$ and Ref.~\cite{daCruzPinhaBarbosa_et_al:2022} for Ba$_2$NaOsO$_6$. The corresponding parameters are listed in \pref{tab:struc_parameters} of \pref{sec:JT}.

Calculations for the magnetically ordered state are performed using a minimal unit cell containing two transition metal cations that allows for a canting of the magnetic moments between the two sublattices according to the experimentally observed magnetic structure shown in \pref{fig:bmrocFM_new}. The simple tetragonal Bravais lattice vectors of this cell correspond to a $1/\sqrt{2} \times 1/\sqrt{2} \times 1$ cell with respect to the conventional cubic cell of the underlying face-centered cubic lattice.
In this case we use a $\Gamma$-centered $8 \times 8 \times 6$ $\vec{k}$-grid, and converge the total energy to a tolerance of at least $10^{-8}$~eV. For the cases where structural relaxations are performed, all force components are converged to values smaller than $10^{-3}$~eV/Å. 

For the paramagnetic case, we use a $2 \times 2 \times 2$ supercell of the basic magnetic cell,  containing 16 $B'$ sites, analogous to what was used in a previous study~\cite{MansouriTehrani/Spaldin:2021}.
We then assign random magnetic moment directions to each transition metal cation, with the additional condition that the total initial magnetization of the cell, calculated assuming the same length for each moment along the chosen direction, is below a tolerance threshold of 0.001~$\mu_B$.
In this case, we use a $3 \times 3 \times 2$ $\Gamma$-centered $\vec{k}$-point grid, and the total energy is converged to an accuracy better than $10^{-5}$~eV. Achieving better convergence is challenging due to the complexity of the noncollinear magnetic configurations, the large supercell, and the two different constraints.
Nevertheless, we verified that the local charge quadrupoles are well converged at this level of precision, and, since the energy scale of interest is almost three orders of magnitude larger than our convergence threshold, a better convergence is not required.

To further ensure that our results are independent from a specific configuration with random magnetic moment directions (given the limited size of our supercell), we always generate four different paramagnetic configurations, and then average over these four calculations to obtain our final results. 
Note that we apply quadrupolar potential shifts of the same magnitude, $|s_t|$, to all transition metal sites in the unit cell with relative signs according to either a ferroic or antiferroic arrangement.
This corresponds to a constraint of the \emph{average} local quadrupole moment, averaged over all sites taking into account the local sign of $s_t$.
We thus obtain the total energy, $E(s_t)$, and the appropriately averaged local quadrupole moment, $\mathcal{Q}_t(s_t)$ for each individual paramagnetic configuration. 
To obtain $E(\mathcal{Q}_t)$, we can then fit an even sixth-order polynomial to all data points, $\{E(s_t), \mathcal{Q}_t(s_t)\}$, obtained for the individual paramagnetic configurations.
Alternatively, we first average all total energies and cell-averaged local quadrupoles for a given $s_t$ over all four paramagnetic configurations and then perform the polynomial fit for $E(\mathcal{Q}_t)$ on these pre-averaged data-points. 
Both procedures lead to almost identical results, with the data presented in \pref{sec:PM-results} corresponding to the latter procedure.

Finally, we have also estimated the effective Hubbard parameter, $U_{\text{eff}}$, for both Ba$_2$MgReO$_6$ and Ba$_2$NaOsO$_6$ using the linear-response-theory-based approach suggested in Ref.~\cite{Cococcioni/deGironcoli:2005}, using finite difference calculations to obtain the site-resolved charge susceptibilities.
For this analysis, we use a $2 \times 2 \times 2$ supercell defined relative to the conventional cubic unit cell of the face-centered cubic lattice (each containing 4 transition metal ions), a $2 \times 2 \times 2$ $\Gamma$-centered $\vec{k}$-grid, and a convergence threshold for the total energy of $10^{-8}$ eV.
We initialize the magnetic moments in a ferromagnetic configuration with magnetization along the [110] direction, and use $U_\text{eff}=0$ for these calculations, i.e., we do not iterate $U_\text{eff}$ to self-consistency. 

We obtain similar values for both materials, \mbox{$U_\text{eff}=1.31$~eV} for Ba$_2$MgReO$_6$ and \mbox{$U_\text{eff}=1.25$~eV} for Ba$_2$NaOsO$_6$. 
These values are comparable to the value $U_\text{eff}=1.8$~eV used in Ref.~\cite{MansouriTehrani/Spaldin:2021} for Ba$_2$MgReO$_6$, which results in a band-gap of $\sim$0.2\,eV, in agreement with the thermal activation energy deduced from electrical resistivity measurements~\cite{Hirai/Hiroi:2019}. In contrast, these values are significantly smaller compared to $U_\text{eff}=3.4$~eV used for Ba$_2$NaOsO$_6$ in Ref.~\cite{FioreMosca_et_al:2021} based on a very simple tight-binding and point-charge estimation in Ref.~\cite{Erickson_et_al:2007}.

In view of the general conceptual uncertainties in the exact definition of the effective Hubbard parameter, and the missing self-consistency with respect to $U_\text{eff}$, we consider our calculated values for $U_\text{eff}$ only as rough but nevertheless very instructive estimates. Thus, in order to allow for a better comparison with Ref.~\cite{MansouriTehrani/Spaldin:2021}, we use $U_\text{eff}=1.8$~eV for both materials, except where otherwise noted.

\section{Results}

\subsection{Emergence of quadrupolar order in the cubic paramagnetic phase}
\label{sec:PM-results}

Our first goal is to establish whether the two materials indeed exhibit a tendency for spontaneous quadrupolar order, even in the absence of dipolar magnetic order and any structural symmetry lowering.

We thus fix the crystal structure to the experimentally observed cubic $Fm\bar{3}m$ structures and simulate the paramagnetic state by using supercells with random orientation of the local magnetic dipole moments and no net magnetization, as described in \pref{sec:comp-details}.
For each paramagnetic configuration, we explore the energy landscape for different quadrupolar orders by applying a constraint on a specific component of the local quadrupoles on the 5$d$ transition metal cations~\cite{Schaufelberger_et_al:2023}, with relative signs on the different sites arranged according to either ferroic or anti-ferroic order. We then vary the strength of the constraint to obtain the energy as function of the size of the average local quadrupole moment.
Specifically, we consider a ferroic ordering of $xy$-quadrupoles (FQ-$xy$ order, symmetry-equivalent to FQ-$xz$ and FQ-$yz$), and an anti-ferroic ordering of $x^2-y^2$-quadrupoles with alternating signs of the quadrupoles along the $z$ direction (AFQ-$x^2-y^2$ order). The latter is consistent with the symmetry-lowering observed in Ba$_2$MgReO$_6$ below $T_q$~\cite{Hirai_et_al:2020, Soh_et_al:2024} and is also the ground state order predicted by the model of Ref.~\cite{Chen/Pereira/Balents:2010}. 
On the other hand, different complementary first principles-based approaches in fact predict the FQ-$xy$ order to be more favorable for Ba$_2$MgReO$_6$ in the absence of any structural distortion~\cite{Soh_et_al:2024, FioreMosca/Franchini/Pourovskii:2024}.

\begin{figure}
    \centering
    \includegraphics[width=0.4\textwidth]{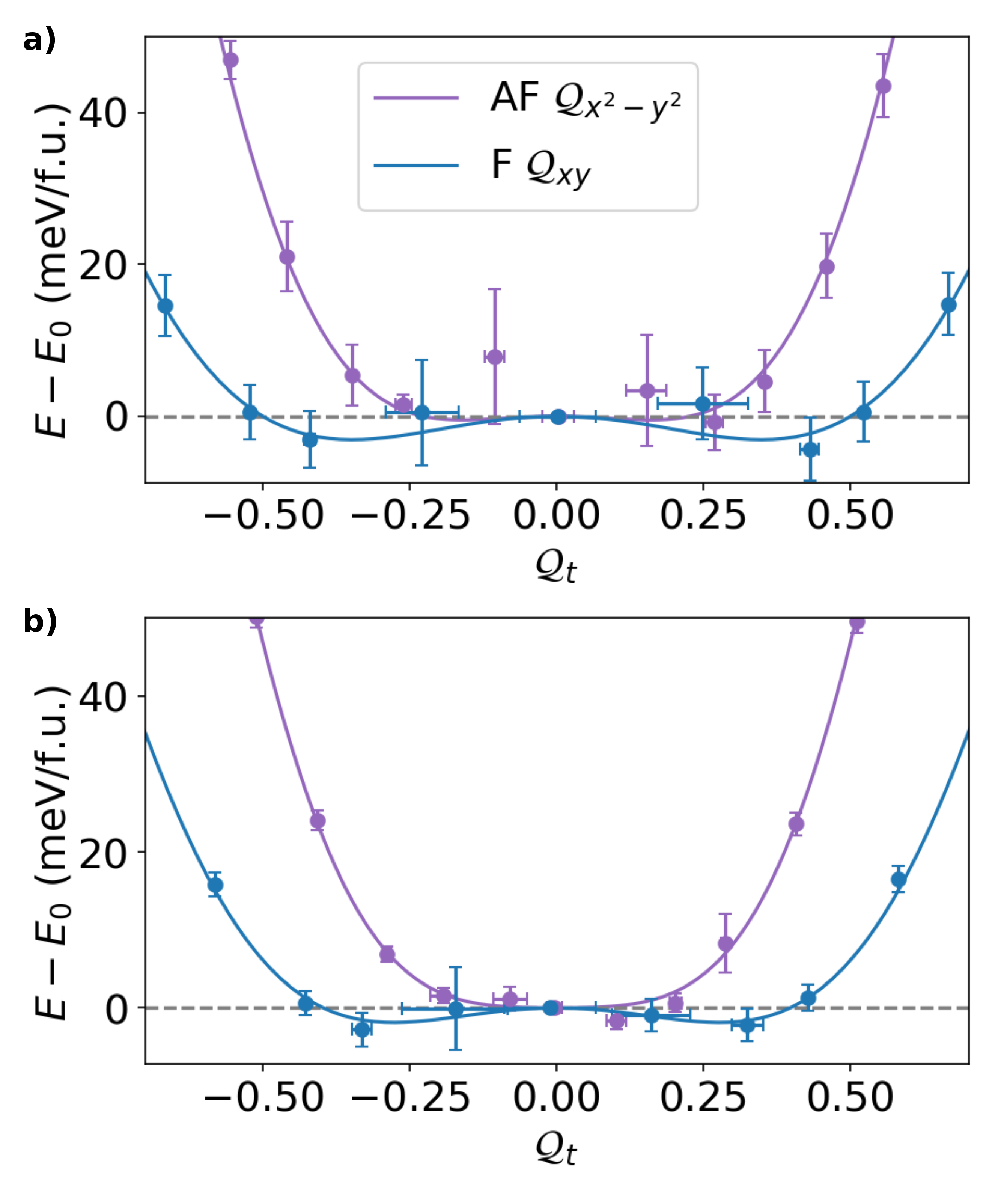}
    \caption{Energy as function of the local quadrupole moment, $Q_t$, for FQ-$xy$ order (blue) and AFQ-$x^2-y^2$ order (purple) for structurally cubic Ba$_2$MgReO$_6$ (a) and Ba$_2$NaOsO$_6$ (b) in the paramagnetic case. Error bars indicate the standard deviation of the energies and local quadrupoles over the different paramagnetic configurations with the same value for the constraint $s_{t}$.}
    \label{fig:bmrobnooPMenergy_even}
\end{figure}

The calculated quadrupolar energy landscapes for both Ba$_2$MgReO$_6$ and Ba$_2$NaOsO$_6$, obtained from appropriate averaging as described in \pref{sec:comp-details}, are shown in \pref{fig:bmrobnooPMenergy_even}.
Both materials exhibit rather similar behavior and, most strikingly, both systems are unstable with respect to a spontaneous FQ-$xy$ type quadrupolar order. The corresponding energy curves exhibit a characteristic ``double well'' shape, with an energy lowering of about 3\,meV/formula unit (f.u.) for Ba$_2$MgReO$_6$ and 2\,meV/f.u. for Ba$_2$NaOsO$_6$ relative to the cubic non-quadrupolar state.
In contrast, the energy associated with the \text{AFQ}-$x^2-y^2$ order exhibits only a very shallow minimum of about 0.5 meV/f.u. for Ba$_2$MgReO$_6$ and no minimum away from $Q_t=0$ for Ba$_2$NaOsO$_6$, even though the energy landscape is notably flat in this region.
We note that both AFQ-$x^2-y^2$ and FQ-$xy$ orders allow for an additional secondary FQ-$3z^2-r^2$ order appearing alongside the imposed quadrupole components. A potential energy-lowering associated with the formation of these $3z^2-r^2$ quadrupoles is thus implicitly included in this data.

Our findings clearly show that, in the absence of any structural distortions or magnetic dipolar order, a FQ-$xy$ order is preferred over an AFQ-$x^2-y^2$ order, in line with previous reports for Ba$_2$MgReO$_6$ using complementary first-principles-based methods~\cite{Soh_et_al:2024, FioreMosca/Franchini/Pourovskii:2024}. This is different to what was obtained from  the purely electronic model presented in Ref.~\cite{Chen/Pereira/Balents:2010}, which incorporated only a single simplified quadrupolar inter-site interaction. 
On the other hand, the energy lowering of 3\,meV/f.u. we obtain for FQ-$xy$ order in Ba$_2$MgReO$_6$ is in very good quantitative agreement with the energy gain obtained from the more general electronic multipolar intersite exchange interactions calculated in Ref.~\cite{FioreMosca/Franchini/Pourovskii:2024}.

The differences between the two materials are somewhat subtle, but overall the tendency for spontaneous quadrupolar order is consistently weaker in Ba$_2$NaOsO$_6$ compared to Ba$_2$MgReO$_6$, represented by the smaller energy gain for FQ-$xy$ order, the smaller magnitude of the corresponding $Q_t$, and the absence of an energy minimum for AFQ-$x^2-y^2$ order in Ba$_2$NaOsO$_6$. 
While these quantitative differences are small, they clearly indicate a weaker tendency for quadrupolar order in Ba$_2$NaOsO$_6$, consistent with the absence of any long-range ordered quadrupolar order or structural Jahn-Teller distortion above $T_m$.

Finally, we note that the relatively large error bars in Fig.~\ref{fig:bmrobnooPMenergy_even}, in particular for small $|Q_t|$, result from large variations of the individual local quadrupole moments in the paramagnetic configurations, which suggests a rather strong influence of the specific local symmetry breaking through the quasi-random orientation of the local magnetic dipole moments. Indeed, if we switch off spin-orbit coupling in these calculations, the variation of the local quadrupoles in the $Q_t\approx 0$ region is significantly reduced. This demonstrates the strong coupling between quadrupole and magnetic dipole moments, mediated by the spin-orbit coupling, which is further analyzed in the next section.

\subsection{Coupling between charge quadrupoles and magnetic dipoles}
\label{subsec:QuadMagcoupling}

Next, we investigate the coupling between the quadrupolar order and the magnetic dipolar order.
For this, we perform analogous calculations of the energy as function of the two relevant quadrupolar order parameters as in the previous section, but now with unconstrained magnetic moments initialized in a ferromagnetic configuration with magnetization along the [110] direction. We chose this configuration because it is the direction along which the magnetization is observed experimentally~\cite{Hirai_et_al:2020, Hirai/Hiroi:2019}. According to our test calculations, it is also the lowest-energy configuration with the magnetic dipole moments confined in the $x$-$y$ plane. 
A magnetization along [110] reduces the magnetic space group symmetry to $Im'm'm$. Nevertheless, we still fix the crystal structure to the experimentally observed perfectly cubic $Fm\bar{3}m$ symmetry, in order to separate the effect of the magnetic symmetry breaking from that of any structural distortion.

\begin{figure}
    \centering
    \includegraphics[width=0.4\textwidth]{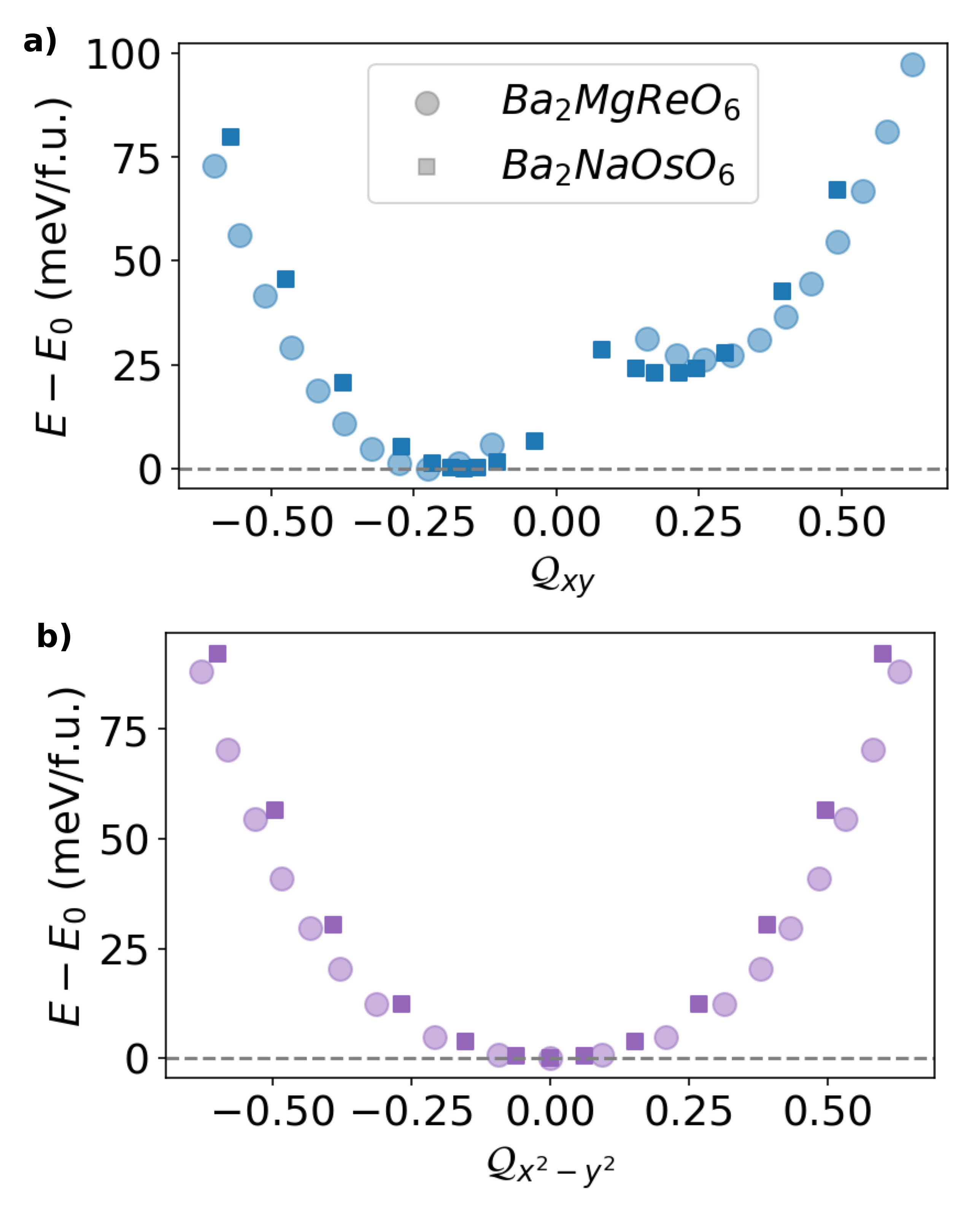}
    \caption{Energy as function of the local quadrupole moments for Ba$_2$MgReO$_6$ (circles) and Ba$_2$NaOsO$_6$ (squares) for the ferromagnetic case with net magnetization along [110] and (a) FQ-$xy$ order or (b) AFQ-$x^2-y^2$ order.}
    \label{fig:FQxyAFQx2y2_energy}
\end{figure}

\pref{fig:FQxyAFQx2y2_energy} (a) depicts the energy landscapes obtained for FQ-$xy$ order for both materials, which exhibit two minima for both positive and negative $Q_{xy} \neq 0$. This resembles to some extent the double-well obtained for the paramagnetic case. However, due to the symmetry reduction resulting from the magnetic order, the two minima for $Q_{xy}\neq0$ are now unequal with the global minimum at $Q_{xy} \approx -0.2$. We note that the significant energy difference between the two minima results solely from the strong spin-orbit coupling.
Specifically, the presence of a magnetization along [110] breaks the fourfold symmetry around the $z$ axis. Consequently, a 90$^\circ$ rotation around $z$, which changes the sign of the $Q_{xy}$ quadrupole, is no longer a symmetry operation, and the states with $\pm Q_{xy}$ become inequivalent. This also means that rotating the magnetization by $90^\circ$ into the $[\bar{1}10]$ direction, switches the two energy minima such that the deeper minimum is now observed for $Q_{xy} \approx +0.2$ (we verified this in our calculations). On the other hand, a reversal of the magnetization by 180$^\circ$ does not affect the quadrupolar energy landscape, since $Q_{xy}$ is invariant under this operation.
The lower energy observed for negative $Q_{xy}$ and magnetization along [110] means that the electrons around the Re and Os cations prefer to accumulate perpendicular to the axis of magnetization. 

We note that the $Im'm'm$ symmetry resulting from the presence of a magnetization along [110] allows for a ferroically ordered $Q_{xy}$ (and $Q_{3z^2-r^2}$) quadrupole moment to emerge without any further symmetry lowering. Thus, the state with $Q_{xy}=0$ does not correspond to an extremum of the total energy, and the minimum for $Q_{xy} \approx -0.2$ corresponds to a calculation with unconstrained quadrupoles.

Fig.~\ref{fig:FQxyAFQx2y2_energy} (b) shows analogous results for the AFQ-$x^2-y^2$ order. It can be seen that emergence of this type of quadrupolar order is now energetically unfavorable for both materials, even though, similar to the paramagnetic case, the energy surface around $Q_{x^2-y^2} \approx 0$ is rather flat, and cannot be fitted well with a simple quadratic parabolic behavior. Note that the sign of the local $Q_{x^2-y^2}$ moments can be reversed, e.g., by a mirror operation perpendicular to [110], which is still a symmetry operation within $Im'm'm$, and thus the energy in \pref{fig:FQxyAFQx2y2_energy}(b) is symmetric with respect to $\pm Q_{x^2-y^2}$.

Another important feature that can be seen from both panels in \pref{fig:FQxyAFQx2y2_energy} is that the tendency toward quadrupole formation appears consistently slightly weaker for Ba$_2$NaOsO$_6$ than for Ba$_2$MgReO$_6$, similar to what is observed for the paramagnetic case in \pref{sec:PM-results}. Specifically, in Fig.~\ref{fig:FQxyAFQx2y2_energy} (a), the energy minimum for Ba$_2$NaOsO$_6$ appears for smaller $|Q_{xy}|$ than for Ba$_2$MgReO$_6$, and in Fig.~\ref{fig:FQxyAFQx2y2_energy} (b), one can recognize a slightly higher stiffness of the energy landscape for Ba$_2$NaOsO$_6$.

\begin{figure}
    \centering 
    \includegraphics[width=0.4\textwidth]{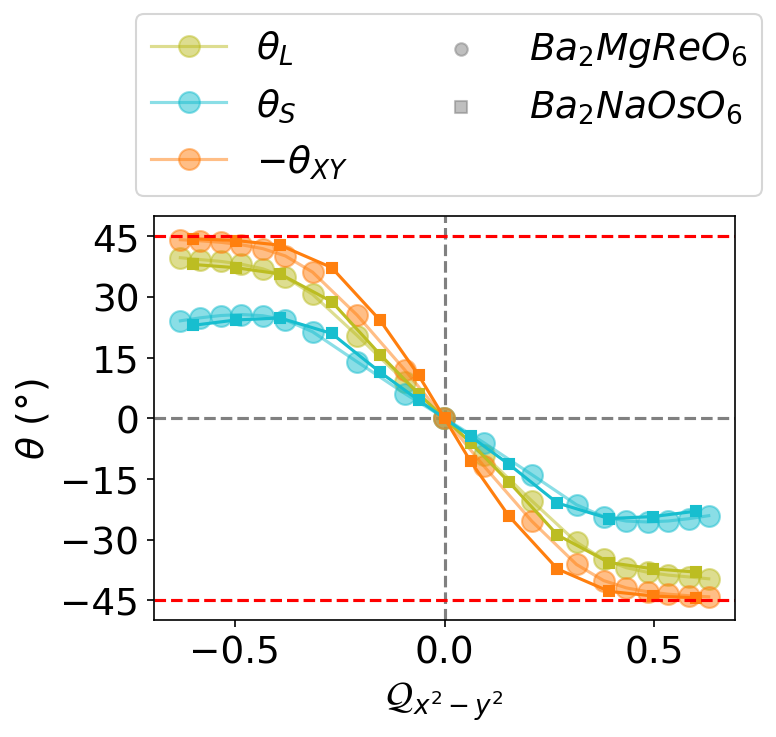}
    \caption{Evolution of canting angles of the local orbital (green) and spin (blue) moments as a function of the local $\mathcal{Q}_{x^2-y^2}$ component for an imposed AFQ-$x^2-y^2$ order, together with the angle $-\theta_{XY}$ (orange) indicating the rotation of the total local quadrupole moment.}
    \label{fig:canting_angle}
\end{figure}

Interestingly, the presence of a nonzero AFQ-$x^2-y^2$ order leads to an antiferromagnetic canting of the Re/Os magnetic moments within the $x$-$y$ plane away from the net magnetization direction, consistent with the canting observed experimentally in the lowest temperature phase of Ba$_2$MgReO$_6$ [{\it cf.} \pref{fig:bmrocFM_new}]. The canting angles $\theta_S$ and $\theta_L$, corresponding to the local spin and orbital moments, respectively, as function of the imposed $Q_{x^2-y^2}$ moment are shown in \pref{fig:canting_angle}.
In addition, together with the increasing local $Q_{x^2-y^2}$ quadrupole, the corresponding $Q_{xy}$ moment is gradually reduced, until it eventually vanishes for $|Q_{x^2-y^2}| \gtrsim 0.5$. This gradual change from a $Q_{xy}$ to a $Q_{x^2-y^2}$ quadrupole, can also be described as a rotation of the local quadrupole by 45$^\circ$ as shown in the following.

Defining quadrupoles $Q_{XY}(\theta)$ and $Q_{X^2-Y^2}(\theta)$ within a local coordinate system $X$-$Y$ that is rotated by an angle $\theta$ around the $z$-axis relative to our global coordinate system, the corresponding transformation of the quadrupole moments can be expressed as: 
\begin{align}\label{eq:quad_rot}
    \begin{pmatrix} Q_{XY}(\theta)\\Q_{X^2-Y^2}(\theta) \end{pmatrix}
    =
    \begin{pmatrix} \mathrm{cos}2\theta & -\mathrm{sin}2\theta \\ \mathrm{sin}2\theta & \mathrm{cos}2\theta\end{pmatrix}
    \begin{pmatrix} Q_{xy}\\Q_{x^2-y^2}\end{pmatrix}.
\end{align}
Thus, the angle defining the orientation of a coordinate system where $Q_{X^2-Y^2}=0$, i.e., describing the rotation of a local $Q_{XY}$ quadrupole, is given by $\theta_{XY}=\frac{1}{2}\mathrm{arctan}\left(-\frac{\mathcal{Q}_{x^2-y^2}}{\mathcal{Q}_{xy}}\right)$. 
In \pref{fig:canting_angle} we plot $-\theta_{XY}$ along with the canting angles of the spin and orbital moment. It can be seen that in particular the canting of the orbital moment closely follows, with opposite sign, the rotation defined by $\-\theta_{XY}$. This makes sense, since the orientation of the local quadrupole reflects the occupation of the local $d$ orbitals, which also sets the local single-ion anisotropy determining the preferred direction of the orbital moment. The observation that the orbital moment rotates opposite to the local quadrupole, means that, while, for the collinear ferromagnetic configuration, the axis of electron accumulation is oriented perpendicular to the magnetization direction, it becomes aligned with the local orbital moment for the maximally canted antiferroic configuration, i.e., for $\theta=\pm 45^\circ$, corresponding to a 90$^\circ$ angle between the two sublattices.
The local spin moment follows the local orbital moment due to the strong spin-orbit coupling, but the fact that the canting of the spin moment is weaker than that of the orbital moment indicates a competition with the inter-site spin-spin exchange favoring a collinear alignment, which leads to an unusual non-collinearity between the local spin and orbital moments. We also note that in these results there is no discernible difference between the two materials.

\begin{figure}
    \centering
    \includegraphics[width=0.4\textwidth]{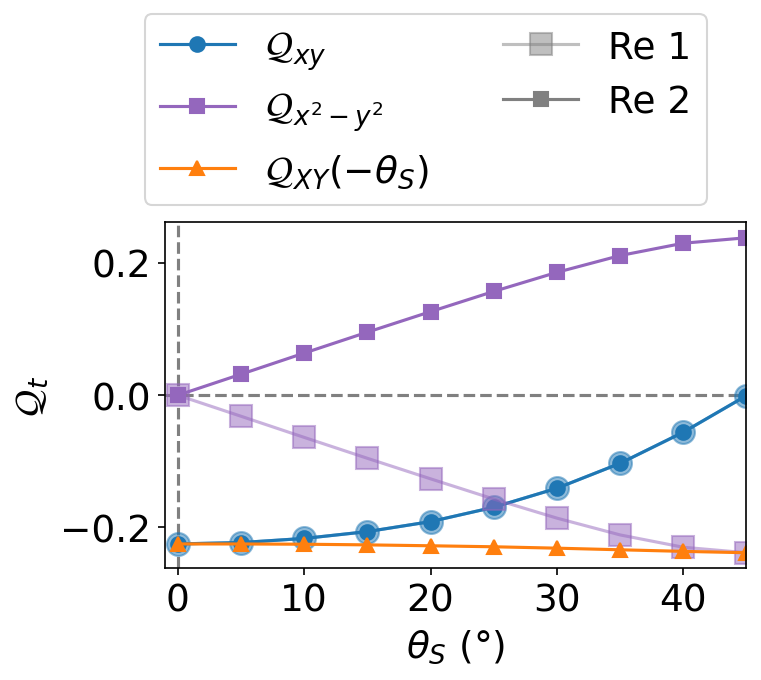}
    \caption{Evolution of the local quadrupole components $Q_{xy}$ and $Q_{x^2-y^2}$ (defined in the global coordinate system) on the two Re sites in the unit cell as a function of the canting angle $\theta_S$ between the spin moments (defined on Re 1) and the [110] direction in Ba$_2$MgReO$_6$. $\mathcal{Q}_{XY}(-\theta_S)$ represents the $xy$-type quadrupole (on Re 1) defined in a local coordinate system that is rotated in the opposite direction as the local spin moment, calculated according to \pref{eq:quad_rot}.}
    \label{fig:quad_vs_canting}
\end{figure}

The same coupling between local magnetic and quadrupole moments can also be observed by constraining the spin moments instead of the quadrupoles. \pref{fig:quad_vs_canting} shows the evolution of the quadrupoles obtained from a set of calculations for Ba$_2$MgReO$_6$ where we constrain the directions of the spin moments to a specific canting angle $\theta_S$.
Increasing $\theta_S$ leads to a gradual suppression of $Q_{xy}$ and the emergence of an increasing AFQ-$x^2-y^2$ component.
We also plot the value of $Q_{XY}(-\theta_S)$, defined according to \pref{eq:quad_rot} in a local coordinate system that rotates opposite to $\theta_S$. Since, to a very good approximation, $Q_{XY}(-\theta_S)$ remains constant for increasing spin canting, this shows that the spin canting results in an essentially rigid rotation of the charge anisotropy around the transition metal sites opposite to that of the spin moment.
We also note that in these calculations the local orbital moments always remain parallel to the local spin moment.
The observation that the local quadrupole moments (and the local orbital moments) remain aligned with the spin moments also indicates that the quadrupolar inter-site coupling, which prefers a FQ-$xy$ alignment, is relatively weak compared to the spin-orbit coupling. 

\subsection{Coupling to the structural Jahn-Teller distortion}
\label{sec:JT}

Up to now, we have only considered purely electronic coupling between the charge multipoles and magnetic dipoles, while the atomic positions were fixed to the perfectly cubic room temperature structures. We have seen that both Ba$_2$MgReO$_6$ and Ba$_2$NaOsO$_6$ exhibit a tendency towards spontaneous quadrupolar order, and that the canting of the local magnetic dipoles away from the [110] direction is related to the AFQ-$x^2-y^2$ order, which, however, is not energetically favorable. We have also seen that the tendency towards quadrupolar order is weaker in Ba$_2$NaOsO$_6$ compared to Ba$_2$MgReO$_6$, but that these quantitative differences are small and that qualitatively both materials behave very similar.

We now explore how the coupling between electronic and structural degrees of freedom affects the quadrupolar and magnetic dipolar order in Ba$_2$MgReO$_6$ and Ba$_2$NaOsO$_6$. We note that previous work, using different first-principles-based computational approaches, have already demonstrated that the coupling to the structural distortion is crucial to stabilize the experimentally observed AFQ-$x^2-y^2$ order in Ba$_2$MgReO$_6$~\cite{Soh_et_al:2024,FioreMosca/Franchini/Pourovskii:2024}.

\begin{table*}
\caption{Experimental and relaxed structural parameters for Ba$_2$MgReO$_6$ and Ba$_2$NaOsO$_6$, together with the local charge quadrupoles, $\mathcal{Q}_t$, the angle $\theta_{XY}$, magnetic spin and orbital dipoles, $\vec{S}$ and $\vec{L}$, absolute value of the total magnetic moment, $|\vec{M}|$, the corresponding canting angles, $\theta_S$, $\theta_L$, $\theta_M$, and the energy differences, $\Delta E$, obtained for the corresponding structures using the specified values for $U_\text{eff}$. The different structures are characterized by their unit cell volumes (per 2 f.u.), the volume ratio of the $B$O$_6$ and $B'$O$_6$ octahedra, $Q_\text{ratio}$, and the (inversion symmetric) distortion modes of the ReO$_6$ octahedra, $Q_1$-$Q_6$, using the Van Vleck notation according to~\cite{Bates:1978} and defined relative to the corresponding cubic $Fm\bar3m$ structure. We also indicate in which cases the spin moments have been constrained to the experimentally reported $\theta_S$ during relaxation.}

\label{tab:struc_parameters}
\centering
\begin{tabular}{|l|ccc|cccc|}
\hline
& \multicolumn{3}{c|}{\textbf{Ba$_2$MgReO$_6$}} & \multicolumn{4}{c|}{\textbf{Ba$_2$NaOsO$_6$}} \\ \hline
Structure & $Fm\bar{3}m$ \cite{Hirai_et_al:2020} & $P4_2/mnm$ \cite{Hirai_et_al:2020} & Relaxed  & $Fm\bar{3}m$ \cite{daCruzPinhaBarbosa_et_al:2022} & Relaxed & Relaxed & Relaxed \\ %\hline
$\theta_S$ constrained? & - & - & No & - & No & No & Yes \\ %\hline
$U_{\text{eff}}$ (eV) & 1.8 & 1.8 & 1.8 & 1.8 & 1.8 & 3.4 & 3.4 \\ %\hline
Vol. (\AA$^3$) & 263.78 & 263.78 & 274.87 & 283.97 & 293.60 & 292.98 & 293.05 \\ %\hline
$Q_{\text{ratio}}$ & 1.322 & 1.323 & 1.305 & 1.723 & 1.715 & 1.733 & 1.734\\ %\hline
$\Delta E$ (meV/f.u.) & 0.146 & 0.148 & 0.000 & 0.118 & 0.000 & 0.000 & 0.008 \\ \hline
\multicolumn{8}{c}{} \\
\multicolumn{8}{c}{\textit{Jahn-Teller modes}} \\ \hline
$Q_1$ (\AA) & 0.000 & -0.001 & 0.076 & 0.000 & 0.055 & 0.043 & 0.043 \\ %\hline
$Q_2$ (\AA) & 0.000 & -0.044 (AF) & -0.036 (AF) & 0.000 & 0.000 & 0.000 & -0.018 (AF) \\ %\hline
$Q_3$ (\AA) & 0.000 & 0.010 & 0.002 & 0.000 & 0.000 & 0.000 & -0.010 \\ %\hline
$Q_4$ (\AA) & 0.000 & 0.000 & 0.000 & 0.000 & -0.012 & 0.014 & 0.000 \\ %\hline
$Q_5$ (\AA) & 0.000 & 0.000 & 0.000 & 0.000 & -0.012 & 0.014 & 0.000 \\ %\hline
$Q_6$ (\AA) & 0.000 & 0.000 & 0.019 & 0.000 & 0.012 & 0.014 & -0.019 \\ \hline
\multicolumn{8}{c}{} \\
\multicolumn{8}{c}{\textit{Charge quadrupoles}} \\ \hline
$\mathcal{Q}_{x^2-y^2}$ & 0.000 & -0.248 & -0.208 (AF) & 0.000 & 0.000 & 0.000 & -0.135 (AF)\\ %\hline
$\mathcal{Q}_{z^2}$ & 0.000 & 0.043 & 0.010 & 0.000 & -0.001 & 0.000 & -0.065 \\ %\hline
$\mathcal{Q}_{yz}$ & -0.126 & 0.000 & 0.001 & -0.088 & 0.091 & -0.111 & 0.000 \\ %\hline
$\mathcal{Q}_{xz}$ & -0.126 & 0.000 & 0.001 & -0.088 & 0.091 & -0.111 & 0.000 \\ %\hline
$\mathcal{Q}_{xy}$ & -0.126 & -0.078 & -0.142 & -0.088 & -0.092 & -0.111 & 0.143 \\ %\hline
$\theta_{XY}$ ($^\circ$) & -  & -36.3 & -27.8 & - & - & - & -68.3 \\ \hline
\multicolumn{8}{c}{} \\
\multicolumn{8}{c}{\textit{Spin moments}} \\ \hline
$S_x$ ($\mu_B$)& -0.449 & -0.283 & -0.341 & -0.375 & -0.379 & -0.415 & 0.245 \\ %\hline
$S_y$ ($\mu_B$)& -0.449 & -0.673 & -0.678 & -0.375 & -0.379 & -0.415 & -0.608\\ %\hline
$S_z$ ($\mu_B$)& -0.449 & 0.001 & 0.003 & -0.375 & 0.376 & -0.416 & 0.000\\ %\hline
$\theta_S$ ($^\circ$) & - & 22.2 & 18.3 & - & - & - & 67.0 \\ \hline
\multicolumn{8}{c}{} \\
\multicolumn{8}{c}{\textit{Orbital moments}} \\ \hline
$L_x$ ($\mu_B$)& 0.151 & 0.050 & 0.085 & 0.114 & 0.119 & 0.140 & -0.095 \\ %\hline
$L_y$ ($\mu_B$)& 0.151 & 0.226 & 0.236 & 0.114 & 0.118 & 0.140 & 0.214\\ %\hline
$L_z$ ($\mu_B$)& 0.151 & 0.000 & -0.001 & 0.115 & -0.118 & 0.140 & 0.000\\ %\hline
$\theta_L$ ($^\circ$) & - & 32.5 & 25.1 & - & - & - & 68.9 \\ \hline
\multicolumn{8}{c}{} \\
\multicolumn{8}{c}{\textit{Total magnetic moments}} \\ \hline
$|\vec{M}|$ ($\mu_B$) & 0.516 & 0.504 & 0.511 & 0.451 & 0.450 & 0.477 & 0.421 \\
$\theta_M$ ($^\circ$) & - & 17.5 & 14.9 & - & - & - & 65.8 \\ \hline
\end{tabular}
\end{table*}

We perform structural relaxations for both materials starting from the reported Jahn-Teller-distorted low temperature structure of Ba$_2$MgReO$_6$~\cite{Hirai_et_al:2020}. \pref{tab:struc_parameters} summarizes the relevant parameters of all relaxed and experimental structures, together with the local charge quadrupoles and the magnetic spin and orbital dipole moments calculated for each structure. The distortion of the oxygen octahedra around Re and Os are characterized by the collective modes, $Q_1$-$Q_6$, using Van Vleck notation according to Ref.~\cite{Bates:1978}, with all modes defined relative to the corresponding cubic $Fm\bar{3}m$ structures. 
We also include data calculated for the experimental cubic structures obtained without confining the magnetic moments to the $x$-$y$ plane. In this case, the lowest energy state corresponds to a collinear ferromagnetic configuration with magnetization along the [111] direction, and a ferroic ordering of $t_{2g}$-type quadrupoles with $\mathcal{Q}_{yz} = \mathcal{Q}_{xz} = \mathcal{Q}_{xy}$. 

Focusing first on the results obtained for Ba$_2$MgReO$_6$, we find that the relaxed structure shows very good agreement with the reported low-temperature experimental structure from~\cite{Hirai_et_al:2020} (and also with the relaxed structure from~\cite{MansouriTehrani/Spaldin:2021}, not shown in \pref{tab:struc_parameters}), except for a slight overestimation of the calculated cell volume typical for the PBE functional. 
Consequently, the calculated quadrupoles and magnetic dipoles are also very similar to those obtained for the experimental $P4/mnm$ structure.
Most importantly, the coupling between electronic and structural degrees of freedom now stabilizes an AFQ-$x^2-y^2$ quadrupolar component and a corresponding $Q_2$ Jahn-Teller distortion, which now becomes energetically more favorable than the ferroic ordering of $t_{2g}$-type quadrupoles in the cubic structure. 
The stabilization of the AFQ-${x^2-y^2}$ order (and the AF-$Q_2$ Jahn-Teller distortion) is also accompanied by an AF canting of the magnetic moments away from the [110] direction, consistent with the analysis of the coupling between charge quadrupoles and magnetic dipoles presented in~\pref{subsec:QuadMagcoupling}.
The $Q_{xy}$ and $Q_{x^2-y^2}$ quadrupoles correspond to a ``quadrupolar angle'', $\theta_{XY}$, of approximately $-28^\circ$ , while orbital and spin moments are canted by about 25$^\circ$ and 18$^\circ$, respectively, relative to the [110] direction, indicating again a competition between the local easy magnetic axis set by the quadrupoles (now energetically stabilized by the Jahn-Teller distortion), the spin-orbit coupling, and the inter-site spin exchange.
This degree of canting is consistent with the previous first-principles study of Ref.~\cite{MansouriTehrani/Spaldin:2021}. Defining the total magnetic moment simply as $\vec{M} = \vec{S} + \vec{L}$ (note that here $\vec{S}$ and $\vec{L}$ are defined in $\mu_\text{B}$), results in a canting angle of about 15$^\circ$ for $\vec{M}$, which is significantly smaller than the experimentally reported value of 40$^\circ$~\cite{Hirai/Hiroi:2019}. However, we note that this angle has merely been estimated from the discrepancy between the observed magnetization of 0.46~$\mu_\text{B}$ per Re and an assumed local magnetic moment of 0.6~$\mu_\text{B}$. Indeed, our calculated $\vec{M}$ corresponds to a net magnetization of 0.49~$\mu_\text{B}$ per Re along [110] and is thus in excellent agreement with the experimentally measured magnetization.

We now turn to the results obtained for Ba$_2$NaOsO$_6$, for which we also perform relaxations using $U_\text{eff}=3.4$\,eV to allow for better comparison with ~\cite{FioreMosca_et_al:2021}. However, we first focus on the calculations with $U_\text{eff}=1.8$\,eV. 
As seen from \pref{tab:struc_parameters}, the initialized AF-$Q_2$ Jahn-Teller distortion is not stable in this case, and the system relaxes to a state with a ferroic ordering of $t_{2g}$-type quadrupoles, a small (ferroic) trigonal distortion of the OsO$_6$ octahedra, corresponding to $Q_4=Q_5=-Q_6 \neq 0$, and a collinear ferromagnetic order with magnetization along $[\bar{1}\bar{1}1]$.
This state is in principle equivalent to the electronic and magnetic ground state obtained for the cubic structure (for both Ba$_2$NaOsO$_6$ and Ba$_2$MgReO$_6$), but with the additional appearance of the $t_{2g}$-type Jahn-Teller modes, consistent with the symmetry lowering stemming from the magnetic and quadrupolar order. (Note that the different magnetization direction only represents a different but equivalent domain state.) 
The resulting trigonal distortion of the OsO$_6$ octahedra also corresponds to the ground state of a $d^1$ ion in an isolated octahedron coupled to $t_{2g}$-like distortion modes, the so-called $t \otimes T$ problem, in the presence of zero or finite spin-orbit coupling~\cite{Bersuker:2006, Streltsov_et_al:2022}.
Finally, we point out that the corresponding unit cell volume and the relative size of the $B$O$_6$ and $B'$O$_6$ octahedra agree very well with the experimental data for the cubic structure, again with the usual PBE overestimation of the total volume. 

The absence of a stable AF-$Q_2$ Jahn-Teller distortion and a resulting canting of the magnetic moments in Ba$_2$NaOsO$_6$ might appear at odds with results of previous first principles calculations presented in Ref.~\cite{FioreMosca_et_al:2021}. However, we note that in the structural relaxations of Ref.~\cite{FioreMosca_et_al:2021}, the spin moments have been constrained to the experimentally reported canting angle of 67$^\circ$. Thus, for a better comparison with Ref~\cite{FioreMosca_et_al:2021}, \pref{tab:struc_parameters} also contains our results of analogous relaxations with constrained spin moments and using $U_\text{eff}=3.4$\,eV. The resulting structure agrees well with that of Ref.~\cite{FioreMosca_et_al:2021}, and one can see that in this case the AF-$Q_2$ distortion indeed remains stable, even though the resulting amplitude is only about half of that obtained for Ba$_2$MgReO$_6$. Furthermore, if the magnetic moments are not constrained during the relaxation, the system relaxes again to a similar state as for $U_\text{eff}=1.8$\,eV, with collinear magnetic moments along [111] and a small ferroic trigonal distortion with $Q_4=Q_5=Q_6$, even though the energy gain compared to the structure with constrained spin moments is rather small. 

These results indicate that the coupling to the Jahn-Teller distortion is significantly weaker in Ba$_2$NaOsO$_6$ compared to Ba$_2$MgReO$_6$. In principle, this appears to be consistent with the absence of any experimentally resolvable Jahn-Teller distortion above $T_m$ in Ba$_2$NaOsO$_6$. Even though our fully relaxed structure exhibits a small trigonal Jahn-Teller distortion, this distortion is rather subtle. The distortion corresponds to oxygen displacements smaller than 0.01\,\AA{}, and the trigonal distortion of the lattice vectors is less than 0.005$^\circ$. Furthermore, the presence of magnetic long-range order in the corresponding calculations might even lead to an overestimation of the corresponding distortion. On the other hand, the absence of a stable AFQ-$x^2-y^2$ order and a corresponding AF-$Q_2$ Jahn-Teller distortion in the relaxed structure also results in an absence of any canting between the magnetic moments, which appears to be in contradiction to experimental observation based on the analysis of NMR spectra~\cite{Lu_et_al:2017}.

\section{Summary and Conclusions}

In this work, we have addressed the possible emergence of quadrupolar order and its coupling to magntic  dipolar order and structural Jahn-Teller distortions for two closely related $5d^1$ double perovskites,  Ba$_2$MgReO$_6$ and  Ba$_2$NaOsO$_6$. We have shown that both materials exhibit a tendency for spontaneous quadrupolar order even in the completely undistorted, perfectly cubic structure, and even in the absence of any magnetic dipolar order. In this case, a ferroic ordering of $t_{2g}$-type quadrupoles, e.g., FQ-$xy$ order, is energetically more favorable than a potential AFQ-$x^2-y^2$ order for both systems, and the overall tendency towards quadrupolar order is slightly weaker for BaNaOsO$_6$ than for Ba$_2$MgReO$_6$. The preference for FQ-$xy$-type order in the absence of any structural distortion is consistent with several previous first-principles-based studies for Ba$_2$MgReO$_6$~\cite{Soh_et_al:2024, FioreMosca/Franchini/Pourovskii:2024} and confirms that a coupling to the structural Jahn-Teller distortion has to be taken into account to explain the experimentally observed low-temperature phase.

For the magnetically ordered case with net magnetization along [110], we have demonstrated an intimate coupling between the local quadrupoles and the local magnetic dipoles. Even though FQ-$xy$ order is still energetically preferred over AFQ-$x^2-y^2$ order in the absence of any structural distortion, the gradual introduction of a nonzero AFQ-$x^2-y^2$ component, equivalent to an alternating rotation of the local charge anisotropy around the $z$-axis, leads to a corresponding canting of the magnetic dipoles away from the net magnetization direction. Thereby, the orbital moments follow the local easy axis set by the quadrupole almost rigidly, while the spin moments ``lag behind'', indicating a competition between the strong spin-orbit coupling, which tends to align the spin and orbital moments, and the usual inter-site spin exchange, which prefers a collinear alignment of spins on neighboring sites. Conversely, an enforced canting of the spin moments also leads to a corresponding alternating rotation of the local quadrupole or, equivalently, an induced AFQ-$x^2-y^2$ component.

We note that an equivalent coupling between the canting of the spin moment and the local charge quadrupoles has also been observed in Ref.~\cite{MansouriTehrani/Spaldin:2021}, albeit calculated for the distorted $P4_2/mnm$ structure. In that case, due to the presence of the corresponding Jahn-Teller distortion, the AFQ-$x^2-y^2$ component persists along the whole range of explored canting angles.
A coupling between quadrupoles and local magnetic dipoles is also consistent with the coupling between the spin canting and local Jahn-Teller distortion observed in first-principles calculations for Ba$_2$NaOsO$_6$ in Ref.~\cite{FioreMosca_et_al:2021}, considering that imposing a Jahn-Teller distortion on the $B'$O$_6$ octahedra will inevitably also generate a corresponding local quadrupole component. 

Finally, considering also small distortions of the cubic structure, we again confirm the energetic stabilization of the AFQ-$x^2-y^2$ by the simultaneous emergence of a corresponding AF-$Q_2$ Jahn-Teller mode for Ba$_2$MgReO$_6$, consistent with the experimentally observed low-temperature phase. In addition, we obtain excellent quantitative agreement with the corresponding experimental data, both for the degree of structural distortion and the net magnetization.
For Ba$_2$NaOsO$_6$, the coupling to the structural distortion is significantly weaker and is not sufficient to stabilize an antiferroic order of local quadrupoles or a corresponding Jahn-Teller distortion. While this is consistent with the absence of any long-range-ordered distorted phase above $T_m$ in Ba$_2$NaOsO$_6$, it appears to be at odds with the reported canted magnetic order below $T_m$, since, according to our analysis (and \cite{MansouriTehrani/Spaldin:2021, FioreMosca_et_al:2021}), such a canting should be tied to an antiferroic quadrupolar order, which, however, appears energetically unfavorable in our calculations. 

Further research is required to resolve this remaining discrepancy for Ba$_2$NaOsO$_6$. We note that, while there is no reason why the PBE+$U$ approach used in our work should be more suitable for Ba$_2$MgReO$_6$ than for Ba$_2$NaOsO$_6$, the small volume overestimation inherent in the PBE functional could in principle introduce some systematic deviations in the corresponding results. 

In summary, by systematically comparing results for two isoelectronic and isostructural 5$d^1$ double perovskites, we have elucidated the mechanisms that govern the emergence of their symmetry-broken low temperature phases with strongly coupled quadrupolar, magnetic dipolar, and structural degrees of freedom. Our work can thus serve as reference for future studies of other $5d^1$ double perovskites, and  also provides a well-defined starting point for future studies relating the observed trends to differences in the underlying structural or electronic parameters (e.g., unit cell volume, $B$-$B'$ size ratio, intersite hopping, degree of $d$-$p$ hybridization, etc.).

\bigskip

\bibliography{BMROvsBNOO}

@article{Agrestini_et_al:2024,
  title = {Origin of {{Magnetism}} in a {{Supposedly Nonmagnetic Osmium Oxide}}},
  author = {Agrestini, S. and Borgatti, F. and Florio, P. and Frassineti, J. and Fiore Mosca, D. and Faure, Q. and Detlefs, B. and Sahle, C. J. and Francoual, S. and Choi, J. and {Garcia-Fernandez}, M. and Zhou, K.-J. and Mitrovi{\'c}, V. F. and Woodward, P. M. and Ghiringhelli, G. and Franchini, C. and Boscherini, F. and Sanna, S. and Moretti Sala, M.},
  year = 2024,
  month = aug,
  journal = {Physical Review Letters},
  volume = {133},
  number = {6},
  pages = {066501},
  issn = {0031-9007, 1079-7114},
  doi = {10.1103/PhysRevLett.133.066501},
  urldate = {2026-02-03},
  langid = {english}
}

@article{Bates:1978,
  title = {Jahn-{{Teller}} Effects in Paramagnetic Crystals},
  author = {Bates, C.A.},
  year = 1978,
  month = jan,
  journal = {Physics Reports},
  volume = {35},
  number = {3},
  pages = {187--304},
  issn = {03701573},
  doi = {10.1016/0370-1573(78)90122-9},
  urldate = {2026-01-15},
  copyright = {https://www.elsevier.com/tdm/userlicense/1.0/},
  langid = {english}
}

@article{Bultmark_et_al:2009,
  title = {Multipole Decomposition of {{LDA}}+{{U}} Energy and Its Application to Actinide Compounds},
  author = {Bultmark, Fredrik and Cricchio, Francesco and Gr{\aa}n{\"a}s, Oscar and Nordstr{\"o}m, Lars},
  year = 2009,
  month = jul,
  journal = {Physical Review B},
  volume = {80},
  number = {3},
  pages = {035121},
  publisher = {American Physical Society},
  doi = {10.1103/PhysRevB.80.035121},
  urldate = {2025-07-15}
}

@article{Chen/Pereira/Balents:2010,
  title = {Exotic Phases Induced by Strong Spin-Orbit Coupling in Ordered Double Perovskites},
  author = {Chen, Gang and Pereira, Rodrigo and Balents, Leon},
  year = 2010,
  month = nov,
  journal = {Physical Review B},
  volume = {82},
  number = {17},
  pages = {174440},
  publisher = {American Physical Society},
  doi = {10.1103/PhysRevB.82.174440},
  urldate = {2026-02-03}
}

@article{Cococcioni/deGironcoli:2005,
  title = {Linear Response Approach to the Calculation of the Effective Interaction Parameters in the {{LDA}}+{{U}} Method},
  author = {Cococcioni, Matteo and {de Gironcoli}, Stefano},
  year = 2005,
  month = jan,
  journal = {Physical Review B},
  volume = {71},
  number = {3},
  pages = {035105},
  publisher = {American Physical Society},
  doi = {10.1103/PhysRevB.71.035105},
  urldate = {2026-02-03}
}

@article{Cong_et_al:2019,
  title = {Evidence from First-Principles Calculations for Orbital Ordering in {$\mathrm{Ba}_2\mathrm{NaOsO}_6$} : {{A Mott}} Insulator with Strong Spin-Orbit Coupling},
  shorttitle = {Evidence from First-Principles Calculations for Orbital Ordering in {{Ba}} 2 {{NaOsO}} 6},
  author = {Cong, R. and Nanguneri, Ravindra and Rubenstein, Brenda and Mitrovi{\'c}, V. F.},
  year = 2019,
  month = dec,
  journal = {Physical Review B},
  volume = {100},
  number = {24},
  pages = {245141},
  issn = {2469-9950, 2469-9969},
  doi = {10.1103/PhysRevB.100.245141},
  urldate = {2026-02-03},
  langid = {english}
}

@article{daCruzPinhaBarbosa_et_al:2022,
  title = {The {{Impact}} of {{Structural Distortions}} on the {{Magnetism}} of {{Double Perovskites Containing}} 5d{$^1$} {{Transition-Metal Ions}}},
  author = {{da Cruz Pinha Barbosa}, Victor and Xiong, Jie and Tran, Phuong Minh and McGuire, Michael A. and Yan, Jiaqiang and Warren, Matthew T. and Aguilar, Rolando Valdes and Zhang, Wenjuan and Randeria, Mohit and Trivedi, Nandini and Haskel, Daniel and Woodward, Patrick M.},
  year = 2022,
  month = feb,
  journal = {Chemistry of Materials},
  volume = {34},
  number = {3},
  pages = {1098--1109},
  publisher = {American Chemical Society},
  issn = {0897-4756},
  doi = {10.1021/acs.chemmater.1c03456},
  urldate = {2026-02-03}
}

@article{Dagotto:1994,
  title = {Correlated Electrons in High-Temperature Superconductors},
  author = {Dagotto, Elbio},
  year = 1994,
  month = jul,
  journal = {Reviews of Modern Physics},
  volume = {66},
  number = {3},
  pages = {763--840},
  publisher = {American Physical Society},
  doi = {10.1103/RevModPhys.66.763},
  urldate = {2026-02-03}
}

@article{Dudarev_et_al:1998,
  title = {Electron-Energy-Loss Spectra and the Structural Stability of Nickel Oxide:  {{An LSDA}}+{{U}} Study},
  shorttitle = {Electron-Energy-Loss Spectra and the Structural Stability of Nickel Oxide},
  author = {Dudarev, S. L. and Botton, G. A. and Savrasov, S. Y. and Humphreys, C. J. and Sutton, A. P.},
  year = 1998,
  month = jan,
  journal = {Physical Review B},
  volume = {57},
  number = {3},
  pages = {1505--1509},
  publisher = {American Physical Society},
  doi = {10.1103/PhysRevB.57.1505},
  urldate = {2026-02-03}
}

@article{Erickson_et_al:2007,
  title = {Ferromagnetism in the {{Mott Insulator}} {$\mathrm{Ba}_2\mathrm{NaOsO}_6$}},
  author = {Erickson, A. S. and Misra, S. and Miller, G. J. and Gupta, R. R. and Schlesinger, Z. and Harrison, W. A. and Kim, J. M. and Fisher, I. R.},
  year = 2007,
  month = jul,
  journal = {Physical Review Letters},
  volume = {99},
  number = {1},
  pages = {016404},
  publisher = {American Physical Society},
  doi = {10.1103/PhysRevLett.99.016404},
  urldate = {2026-02-03}
}

@article{FioreMosca_et_al:2021,
  title = {Interplay between Multipolar Spin Interactions, {{Jahn-Teller}} Effect, and Electronic Correlation in a {$J_{\mathrm{eff}}=3/2$} Insulator},
  author = {Fiore Mosca, Dario and Pourovskii, Leonid V. and Kim, Beom Hyun and Liu, Peitao and Sanna, Samuele and Boscherini, Federico and Khmelevskyi, Sergii and Franchini, Cesare},
  year = 2021,
  month = mar,
  journal = {Physical Review B},
  volume = {103},
  number = {10},
  pages = {104401},
  issn = {2469-9950, 2469-9969},
  doi = {10.1103/PhysRevB.103.104401},
  urldate = {2026-02-03},
  langid = {english}
}

@article{FioreMosca_et_al:2024,
  title = {The {{Mott}} Transition in the 5d{$^1$} Compound {$\mathrm{Ba}_2\mathrm{MgReO}_6$} : {{A DFT}}+{{DMFT}} Study with {{PAW}} Spinor Projectors},
  shorttitle = {The {{Mott}} Transition in the 5d 1 Compound {{Ba}} 2 {{NaOsO}} 6},
  author = {Fiore Mosca, Dario and Schnait, Hermann and Celiberti, Lorenzo and Aichhorn, Markus and Franchini, Cesare},
  year = 2024,
  month = jan,
  journal = {Computational Materials Science},
  volume = {233},
  pages = {112764},
  issn = {09270256},
  doi = {10.1016/j.commatsci.2023.112764},
  urldate = {2026-02-03},
  langid = {english}
}

@article{FioreMosca/Franchini/Pourovskii:2024,
  title = {Interplay of Superexchange and Vibronic Effects in the Hidden Order of {$\mathrm{Ba}_2\mathrm{MgReO}_6$} from First Principles},
  author = {Fiore Mosca, Dario and Franchini, Cesare and Pourovskii, Leonid V.},
  year = 2024,
  month = nov,
  journal = {Physical Review B},
  volume = {110},
  number = {20},
  pages = {L201101},
  publisher = {American Physical Society},
  doi = {10.1103/PhysRevB.110.L201101},
  urldate = {2026-02-03}
}

@article{Gangopadhyay/Pickett:2015,
  title = {Spin-Orbit Coupling, Strong Correlation, and Insulator-Metal Transitions: The {$J_{\mathrm{eff}}=3/2$} Ferromagnetic {{Dirac-Mott}} Insulator {$\mathrm{Ba}_2\mathrm{NaOsO}_6$}},
  shorttitle = {Spin-Orbit Coupling, Strong Correlation, and Insulator-Metal Transitions},
  author = {Gangopadhyay, Shruba and Pickett, Warren E.},
  year = 2015,
  month = jan,
  journal = {Physical Review B},
  volume = {91},
  number = {4},
  pages = {045133},
  issn = {1098-0121, 1550-235X},
  doi = {10.1103/PhysRevB.91.045133},
  urldate = {2026-02-03},
  copyright = {http://link.aps.org/licenses/aps-default-license},
  langid = {english}
}

@incollection{Hafner/Kresse:1997,
  title = {The {{Vienna AB-Initio Simulation Program VASP}}: {{An Efficient}} and {{Versatile Tool}} for {{Studying}} the {{Structural}}, {{Dynamic}}, and {{Electronic Properties}} of {{Materials}}},
  shorttitle = {The {{Vienna AB-Initio Simulation Program VASP}}},
  booktitle = {Properties of {{Complex Inorganic Solids}}},
  author = {Hafner, J. and Kresse, G.},
  editor = {Gonis, Antonios and Meike, Annemarie and Turchi, Patrice E. A.},
  year = 1997,
  pages = {69--82},
  publisher = {Springer US},
  address = {Boston, MA},
  doi = {10.1007/978-1-4615-5943-6_10},
  urldate = {2026-02-03},
  isbn = {978-1-4615-5943-6},
  langid = {english}
}

@article{Hirai_et_al:2020,
  title = {Detection of Multipolar Orders in the Spin-Orbit-Coupled 5d {{Mott}} Insulator {$\mathrm{Ba}_2\mathrm{MgReO}_6$}},
  author = {Hirai, Daigorou and Sagayama, Hajime and Gao, Shang and Ohsumi, Hiroyuki and Chen, Gang and Arima, Taka-hisa and Hiroi, Zenji},
  year = 2020,
  month = jun,
  journal = {Physical Review Research},
  volume = {2},
  number = {2},
  pages = {022063},
  publisher = {American Physical Society},
  doi = {10.1103/PhysRevResearch.2.022063},
  urldate = {2026-02-03}
}

@article{Hirai/Hiroi:2019,
  title = {Successive {{Symmetry Breaking}} in a {$J_{\mathrm{eff}}=3/2$} {{Quartet}} in the {{Spin}}--{{Orbit Coupled Insulator}}{$\mathrm{Ba}_2\mathrm{MgReO}_6$}},
  author = {Hirai, Daigorou and Hiroi, Zenji},
  year = 2019,
  month = jun,
  journal = {Journal of the Physical Society of Japan},
  volume = {88},
  number = {6},
  pages = {064712},
  publisher = {The Physical Society of Japan},
  issn = {0031-9015},
  doi = {10.7566/JPSJ.88.064712},
  urldate = {2024-11-28}
}

@article{Imada/Fujimori/Tokura:1998,
  title = {Metal-Insulator Transitions},
  author = {Imada, Masatoshi and Fujimori, Atsushi and Tokura, Yoshinori},
  year = 1998,
  month = oct,
  journal = {Reviews of Modern Physics},
  volume = {70},
  number = {4},
  pages = {1039--1263},
  publisher = {American Physical Society},
  doi = {10.1103/RevModPhys.70.1039},
  urldate = {2026-02-03}
}

@article{Ishizuka/Balents:2014,
  title = {Magnetism in ${{S}}=\frac{1}{2}$ Double Perovskites with Strong Spin-Orbit Interactions},
  author = {Ishizuka, Hiroaki and Balents, Leon},
  year = 2014,
  month = nov,
  journal = {Physical Review B},
  volume = {90},
  number = {18},
  pages = {184422},
  publisher = {American Physical Society},
  doi = {10.1103/PhysRevB.90.184422},
  urldate = {2026-02-03}
}

@article{Iwahara/Chibotaru:2023,
  title = {Vibronic Order and Emergent Magnetism in Cubic d{$^1$} Double Perovskites},
  author = {Iwahara, Naoya and Chibotaru, Liviu F.},
  year = 2023,
  month = jun,
  journal = {Physical Review B},
  volume = {107},
  number = {22},
  pages = {L220404},
  publisher = {American Physical Society},
  doi = {10.1103/PhysRevB.107.L220404},
  urldate = {2026-02-03}
}

@article{Kresse/Joubert:1999,
  title = {From Ultrasoft Pseudopotentials to the Projector Augmented-Wave Method},
  author = {Kresse, G. and Joubert, D.},
  year = 1999,
  month = jan,
  journal = {Physical Review B},
  volume = {59},
  number = {3},
  pages = {1758--1775},
  publisher = {American Physical Society},
  doi = {10.1103/PhysRevB.59.1758},
  urldate = {2026-02-03}
}

@article{Liu_et_al:2018,
  title = {Nature of Lattice Distortions in the Cubic Double Perovskite {$\mathrm{Ba}_2\mathrm{NaOsO}_6$}},
  author = {Liu, W. and Cong, R. and Reyes, A. P. and Fisher, I. R. and Mitrovi{\'c}, V. F.},
  year = 2018,
  month = jun,
  journal = {Physical Review B},
  volume = {97},
  number = {22},
  pages = {224103},
  issn = {2469-9950, 2469-9969},
  doi = {10.1103/PhysRevB.97.224103},
  urldate = {2026-02-03},
  langid = {english}
}

@article{Liu_et_al:2018a,
  title = {Phase Diagram of {$\mathrm{Ba}_2\mathrm{NaOsO}_6$}, a {{Mott}} Insulator with Strong Spin Orbit Interactions},
  author = {Liu, W. and Cong, R. and Garcia, E. and Reyes, A.P. and Lee, H.O. and Fisher, I.R. and Mitrovi{\'c}, V.F.},
  year = 2018,
  month = may,
  journal = {Physica B: Condensed Matter},
  volume = {536},
  pages = {863--866},
  issn = {09214526},
  doi = {10.1016/j.physb.2017.08.062},
  urldate = {2026-02-03},
  langid = {english}
}

@article{Lu_et_al:2017,
  title = {Magnetism and Local Symmetry Breaking in a {{Mott}} Insulator with Strong Spin Orbit Interactions},
  author = {Lu, L. and Song, M. and Liu, W. and Reyes, A. P. and Kuhns, P. and Lee, H. O. and Fisher, I. R. and Mitrovi{\'c}, V. F.},
  year = 2017,
  month = feb,
  journal = {Nature Communications},
  volume = {8},
  number = {1},
  pages = {14407},
  publisher = {Nature Publishing Group},
  issn = {2041-1723},
  doi = {10.1038/ncomms14407},
  urldate = {2026-02-03},
  copyright = {2017 The Author(s)},
  langid = {english}
}

@article{Ma/Dudarev:2015,
  title = {Constrained Density Functional for Noncollinear Magnetism},
  author = {Ma, Pui-Wai and Dudarev, S. L.},
  year = 2015,
  month = feb,
  journal = {Physical Review B},
  volume = {91},
  number = {5},
  pages = {054420},
  publisher = {American Physical Society},
  doi = {10.1103/PhysRevB.91.054420},
  urldate = {2026-02-03}
}

@article{Maharaj_et_al:2020a,
  title = {Octupolar versus {{Neel Order}} in {{Cubic}} d{$^2$} {{Double Perovskites}}},
  author = {Maharaj, D. D. and Sala, G. and Stone, M. B. and Kermarrec, E. and Ritter, C. and Fauth, F. and Marjerrison, C. A. and Greedan, J. E. and Paramekanti, A. and Gaulin, B. D.},
  year = 2020,
  month = feb,
  journal = {Physical Review Letters},
  volume = {124},
  number = {8},
  pages = {087206},
  publisher = {American Physical Society},
  doi = {10.1103/PhysRevLett.124.087206},
  urldate = {2026-02-11}
}

@article{MansouriTehrani/Spaldin:2021,
  title = {Untangling the Structural, Magnetic Dipole, and Charge Multipolar Orders in {$\mathrm{Ba}_2\mathrm{MgReO}_6$}},
  author = {Mansouri Tehrani, Aria and Spaldin, Nicola A.},
  year = 2021,
  month = oct,
  journal = {Physical Review Materials},
  volume = {5},
  number = {10},
  pages = {104410},
  issn = {2475-9953},
  doi = {10.1103/PhysRevMaterials.5.104410},
  urldate = {2026-02-03},
  langid = {english}
}

@article{Marjerrison_et_al:2016,
  title = {Cubic {$\mathrm{Re}^{6+}$} (5$d^1$) Double Perovskites $\mathrm{Ba}_2\mathrm{MgReO}_6$, $\mathrm{Ba}_2\mathrm{ZnReO}_6$, and $\mathrm{Ba}_2\mathrm{Y}_{2/3}\mathrm{ReO}_6$: Magnetism, Heat Capacity, $\mu$SR, and Neutron Scattering Studies and Comparison with Theory},
  shorttitle = {Cubic {{Re6}}+ (5d1) {{Double Perovskites}}, {{Ba2MgReO6}}, {{Ba2ZnReO6}}, and {{Ba2Y2}}/{{3ReO6}}},
  author = {Marjerrison, Casey A. and Thompson, Corey M. and Sala, Gabrielle and Maharaj, Dalini D. and Kermarrec, Edwin and Cai, Yipeng and Hallas, Alannah M. and Wilson, Murray N. and Munsie, Timothy J. S. and Granroth, Garrett E. and Flacau, Roxana and Greedan, John E. and Gaulin, Bruce D. and Luke, Graeme M.},
  year = 2016,
  month = oct,
  journal = {Inorganic Chemistry},
  volume = {55},
  number = {20},
  pages = {10701--10713},
  publisher = {American Chemical Society},
  issn = {0020-1669},
  doi = {10.1021/acs.inorgchem.6b01933},
  urldate = {2024-12-02}
}

@article{Martins/Aichhorn/Biermann:2017,
  title = {Coulomb Correlations in 4d and 5d Oxides from First Principles---or How Spin--Orbit Materials Choose Their Effective Orbital Degeneracies},
  author = {Martins, C and Aichhorn, M and Biermann, S},
  year = 2017,
  month = may,
  journal = {Journal of Physics: Condensed Matter},
  volume = {29},
  number = {26},
  pages = {263001},
  publisher = {IOP Publishing},
  issn = {0953-8984},
  doi = {10.1088/1361-648X/aa648f},
  urldate = {2026-02-03},
  langid = {english}
}

@article{Merkel/Tehrani/Ederer:2024,
  title = {Probing the {{Mott}} Insulating Behavior of {$\mathrm{Ba}_2\mathrm{MgReO}_6$} with {{DFT}} + {{DMFT}}},
  author = {Merkel, Maximilian E. and Tehrani, Aria Mansouri and Ederer, Claude},
  year = 2024,
  month = jun,
  journal = {Physical Review Research},
  volume = {6},
  number = {2},
  pages = {023233},
  issn = {2643-1564},
  doi = {10.1103/PhysRevResearch.6.023233},
  urldate = {2026-02-03},
  langid = {english}
}

@article{Otsuki_et_al:2024,
  title = {Multipolar Ordering from Dynamical Mean Field Theory with Application to {$\mathrm{CeB}_{6}$}},
  author = {Otsuki, Junya and Yoshimi, Kazuyoshi and Shinaoka, Hiroshi and Jeschke, Harald O.},
  year = 2024,
  month = jul,
  journal = {Physical Review B},
  volume = {110},
  number = {3},
  pages = {035104},
  publisher = {American Physical Society},
  doi = {10.1103/PhysRevB.110.035104},
  urldate = {2026-02-03}
}

@article{Perdew/Burke/Ernzerhof:1996,
  title = {Generalized {{Gradient Approximation Made Simple}}},
  author = {Perdew, John P. and Burke, Kieron and Ernzerhof, Matthias},
  year = 1996,
  month = oct,
  journal = {Physical Review Letters},
  volume = {77},
  number = {18},
  pages = {3865--3868},
  publisher = {American Physical Society},
  doi = {10.1103/PhysRevLett.77.3865},
  urldate = {2026-02-03}
}

@article{Rau/Lee/Kee:2016,
  title = {Spin-{{Orbit Physics Giving Rise}} to {{Novel Phases}} in {{Correlated Systems}}: {{Iridates}} and {{Related Materials}}},
  shorttitle = {Spin-{{Orbit Physics Giving Rise}} to {{Novel Phases}} in {{Correlated Systems}}},
  author = {Rau, Jeffrey G. and Lee, Eric Kin-Ho and Kee, Hae-Young},
  year = 2016,
  month = mar,
  journal = {Annual Review of Condensed Matter Physics},
  volume = {7},
  number = {Volume 7, 2016},
  pages = {195--221},
  publisher = {Annual Reviews},
  issn = {1947-5454, 1947-5462},
  doi = {10.1146/annurev-conmatphys-031115-011319},
  urldate = {2026-02-03},
  langid = {english}
}

@article{Romhanyi/Balents/Jackeli:2017,
  title = {Spin-{{Orbit Dimers}} and {{Noncollinear Phases}} in d$^1$ {{Cubic Double Perovskites}}},
  author = {Romh{\'a}nyi, Judit and Balents, Leon and Jackeli, George},
  year = 2017,
  month = may,
  journal = {Physical Review Letters},
  volume = {118},
  number = {21},
  pages = {217202},
  publisher = {American Physical Society},
  doi = {10.1103/PhysRevLett.118.217202},
  urldate = {2026-02-03}
}

@article{Santini_et_al:2009,
  title = {Multipolar Interactions in f-Electron Systems: {{The}} Paradigm of Actinide Dioxides},
  shorttitle = {Multipolar Interactions in f-Electron Systems},
  author = {Santini, Paolo and Carretta, Stefano and Amoretti, Giuseppe and Caciuffo, Roberto and Magnani, Nicola and Lander, Gerard H.},
  year = 2009,
  month = jun,
  journal = {Reviews of Modern Physics},
  volume = {81},
  number = {2},
  pages = {807--863},
  publisher = {American Physical Society},
  doi = {10.1103/RevModPhys.81.807},
  urldate = {2025-03-27}
}

@article{Savary/Balents:2016,
  title = {Quantum Spin Liquids: A Review},
  shorttitle = {Quantum Spin Liquids},
  author = {Savary, Lucile and Balents, Leon},
  year = 2016,
  month = nov,
  journal = {Reports on Progress in Physics},
  volume = {80},
  number = {1},
  pages = {016502},
  publisher = {IOP Publishing},
  issn = {0034-4885},
  doi = {10.1088/0034-4885/80/1/016502},
  urldate = {2026-02-03},
  langid = {english}
}

@article{Schaufelberger_et_al:2023,
  title = {Exploring Energy Landscapes of Charge Multipoles Using Constrained Density Functional Theory},
  author = {Schaufelberger, Luca and Merkel, Maximilian E. and Tehrani, Aria Mansouri and Spaldin, Nicola A. and Ederer, Claude},
  year = 2023,
  month = sep,
  journal = {Physical Review Research},
  volume = {5},
  number = {3},
  pages = {033172},
  publisher = {American Physical Society},
  doi = {10.1103/PhysRevResearch.5.033172},
  urldate = {2026-02-03}
}

@article{Soh_et_al:2024,
  title = {Spectroscopic Signatures and Origin of Hidden Order in {{$\mathrm{Ba}_2\mathrm{MgReO}_6$}}},
  author = {Soh, Jian-Rui and Merkel, Maximilian E. and Pourovskii, Leonid V. and {\v Z}ivkovi{\'c}, Ivica and Malanyuk, Oleg and P{\'a}sztorov{\'a}, Jana and Francoual, Sonia and Hirai, Daigorou and Urru, Andrea and Tolj, Davor and Fiore Mosca, Dario and Yazyev, Oleg V. and Spaldin, Nicola A. and Ederer, Claude and R{\o}nnow, Henrik M.},
  year = 2024,
  month = nov,
  journal = {Nature Communications},
  volume = {15},
  number = {1},
  pages = {10383},
  issn = {2041-1723},
  doi = {10.1038/s41467-024-53893-z},
  urldate = {2026-02-03},
  langid = {english}
}

@article{Streltsov_et_al:2022,
  title = {Interplay of the {{Jahn-Teller}} Effect and Spin-Orbit Coupling: {{The}} Case of Trigonal Vibrations},
  shorttitle = {Interplay of the {{Jahn-Teller}} Effect and Spin-Orbit Coupling},
  author = {Streltsov, Sergey V. and Temnikov, Fedor V. and Kugel, Kliment I. and Khomskii, Daniel I.},
  year = 2022,
  month = may,
  journal = {Physical Review B},
  volume = {105},
  number = {20},
  pages = {205142},
  publisher = {American Physical Society},
  doi = {10.1103/PhysRevB.105.205142},
  urldate = {2026-01-15}
}

@article{Suzuki/Ikeda/Oppeneer:2018,
  title = {First-Principles {{Theory}} of {{Magnetic Multipoles}} in {{Condensed Matter Systems}}},
  author = {Suzuki, Michi-To and Ikeda, Hiroaki and Oppeneer, Peter M.},
  year = 2018,
  month = apr,
  journal = {Journal of the Physical Society of Japan},
  volume = {87},
  number = {4},
  pages = {041008},
  publisher = {The Physical Society of Japan},
  issn = {0031-9015},
  doi = {10.7566/JPSJ.87.041008},
  urldate = {2025-03-21}
}

@article{Takayama_et_al:2021,
  title = {Spin--{{Orbit-Entangled Electronic Phases}} in 4d and 5d {{Transition-Metal Compounds}}},
  author = {Takayama, Tomohiro and Chaloupka, Ji{\v r}{\'i} and Smerald, Andrew and Khaliullin, Giniyat and Takagi, Hidenori},
  year = 2021,
  month = jun,
  journal = {Journal of the Physical Society of Japan},
  volume = {90},
  number = {6},
  pages = {062001},
  publisher = {The Physical Society of Japan},
  issn = {0031-9015},
  doi = {10.7566/JPSJ.90.062001},
  urldate = {2024-11-18}
}

@article{Urru_et_al:2023a,
  title = {Neutron Scattering from Local Magnetoelectric Multipoles: {{A}} Combined Theoretical, Computational, and Experimental Perspective},
  shorttitle = {Neutron Scattering from Local Magnetoelectric Multipoles},
  author = {Urru, Andrea and Soh, Jian-Rui and Qureshi, Navid and Stunault, Anne and Roessli, Bertrand and R{\o}nnow, Henrik M. and Spaldin, Nicola A.},
  year = 2023,
  month = sep,
  journal = {Physical Review Research},
  volume = {5},
  number = {3},
  pages = {033147},
  publisher = {American Physical Society},
  doi = {10.1103/PhysRevResearch.5.033147},
  urldate = {2026-02-11}
}

@article{Vasala/Karppinen:2015,
  title = {{$\mathrm{A}_2\mathrm{B'B''O}_6$} Perovskites: A Review},
  shorttitle = {{\emph{A}}2{{{\emph{B}}}}{$\prime$}{{{\emph{B}}}}{${''}$}{{O6}} Perovskites},
  author = {Vasala, Sami and Karppinen, Maarit},
  year = 2015,
  month = may,
  journal = {Progress in Solid State Chemistry},
  volume = {43},
  number = {1},
  pages = {1--36},
  issn = {0079-6786},
  doi = {10.1016/j.progsolidstchem.2014.08.001},
  urldate = {2024-12-02}
}

@article{Verbeek/Urru/Spaldin:2023,
  title = {Hidden Orders and (Anti-)Magnetoelectric Effects in {$\mathrm{Cr}_2\mathrm{O}_3$} and {$\alpha-\mathrm{Fe}_2\mathrm{O}_3$}},
  author = {Verbeek, Xanthe H. and Urru, Andrea and Spaldin, Nicola A.},
  year = 2023,
  month = nov,
  journal = {Physical Review Research},
  volume = {5},
  number = {4},
  pages = {L042018},
  publisher = {American Physical Society},
  doi = {10.1103/PhysRevResearch.5.L042018},
  urldate = {2026-02-11}
}

@article{Willa_et_al:2019,
  title = {Phase Transition Preceding Magnetic Long-Range Order in the Double Perovskite {$\mathrm{Ba}_2\mathrm{NaOsO}_6$}},
  author = {Willa, Kristin and Willa, Roland and Welp, Ulrich and Fisher, Ian R. and Rydh, Andreas and Kwok, Wai-Kwong and Islam, Zahir},
  year = 2019,
  month = jul,
  journal = {Physical Review B},
  volume = {100},
  number = {4},
  pages = {041108},
  issn = {2469-9950, 2469-9969},
  doi = {10.1103/PhysRevB.100.041108},
  urldate = {2026-02-03},
  langid = {english}
}

@article{Witczak-Krempa_et_al:2014,
  title = {Correlated {{Quantum Phenomena}} in the {{Strong Spin-Orbit Regime}}},
  author = {{Witczak-Krempa}, William and Chen, Gang and Kim, Yong Baek and Balents, Leon},
  year = 2014,
  month = mar,
  journal = {Annual Review of Condensed Matter Physics},
  volume = {5},
  number = {Volume 5, 2014},
  pages = {57--82},
  publisher = {Annual Reviews},
  issn = {1947-5454, 1947-5462},
  doi = {10.1146/annurev-conmatphys-020911-125138},
  urldate = {2025-03-21},
  langid = {english}
}

@article{Xiang/Whangbo:2007,
  title = {Cooperative Effect of Electron Correlation and Spin-Orbit Coupling on the Electronic and Magnetic Properties of {$\mathrm{Ba}_2\mathrm{NaOsO}_6$}},
  author = {Xiang, H. J. and Whangbo, M.-H.},
  year = 2007,
  month = feb,
  journal = {Physical Review B},
  volume = {75},
  number = {5},
  pages = {052407},
  issn = {1098-0121, 1550-235X},
  doi = {10.1103/PhysRevB.75.052407},
  urldate = {2026-02-03},
  copyright = {http://link.aps.org/licenses/aps-default-license},
  langid = {english}
}

@article{Zivkovic_et_al:2024,
  title = {Dynamic {{Jahn-Teller}} Effect in the Strong Spin-Orbit Coupling Regime},
  author = {{\v Z}ivkovi{\'c}, Ivica and Soh, Jian-Rui and Malanyuk, Oleg and Yadav, Ravi and Pisani, Federico and Tehrani, Aria M. and Tolj, Davor and Pasztorova, Jana and Hirai, Daigorou and Wei, Yuan and Zhang, Wenliang and Galdino, Carlos and Yu, Tianlun and Ishii, Kenji and Demuer, Albin and Yazyev, Oleg V. and Schmitt, Thorsten and R{\o}nnow, Henrik M.},
  year = 2024,
  month = oct,
  journal = {Nature Communications},
  volume = {15},
  number = {1},
  pages = {8587},
  publisher = {Nature Publishing Group},
  issn = {2041-1723},
  doi = {10.1038/s41467-024-52935-w},
  urldate = {2026-02-03},
  copyright = {2024 The Author(s)},
  langid = {english}
}

@book{Bersuker:2006,
  title = {The {{Jahn-Teller Effect}}},
  author = {Bersuker, Isaac},
  year = 2006,
  publisher = {Cambridge University Press},
  address = {Cambridge},
  doi = {10.1017/CBO9780511524769},
  urldate = {2026-02-17}
}

\end{document}